\documentclass[pre,amsmath,twocolumn]{revtex4-1}
\usepackage[normalem]{ulem}
\usepackage{epsfig}
\usepackage{amsfonts} 
\usepackage{amsmath} 
\usepackage{nicefrac}
\usepackage{color} 
\usepackage[pdfencoding=auto, psdextra]{hyperref}
\setcitestyle{square}
\usepackage{subfigure} 
\usepackage{natbib} 
\usepackage{graphicx,amssymb} 

\newcommand{\red}{\textcolor{red}}
\newcommand{\blue}{\textcolor{blue}}
\newcommand{\ignore}[1]{} 

\def\eq#1{${#1}$} 
\def\EQ#1#2{\begin{equation}{#1}\label{#2}\end{equation}}

\newcommand{\Intd}{\textrm{d}}

\usepackage{mathrsfs}
\usepackage{commath}

\usepackage{gensymb} 

\setcitestyle{square}
\usepackage{subfigure}

\usepackage{enumitem}
 \usepackage{wasysym} 

\begin{document}

\title{
{Infinite ergodic theory meets Boltzmann statistics}} 

\author{Erez Aghion\textsuperscript{a,b,c},  David A. Kessler\textsuperscript{a}, Eli Barkai\textsuperscript{a,b}}
 \affiliation{\makebox[\textwidth][c]{a) Department of Physics, b) Institute of Nanotechnology and Advanced Materials, Bar-Ilan University, Ramat-Gan 52900, Israel}\\ c) Max-Planck Institute for the Physics of Complex Systems, Dresden D-01187, Germany } 

\begin{abstract}

  We investigate  the overdamped stochastic dynamics of a particle in an asymptotically flat external potential field, in contact with a thermal bath. For an infinite system size, the particles may escape the force field and diffuse freely at large length scales.  The partition function diverges and hence the standard  canonical ensemble  fails. This is replaced
with tools stemming from infinite ergodic theory. Boltzmann-Gibbs statistics, even though not normalized, still describes integrable observables, like energy and occupation times.  The Boltzmann infinite density is derived heuristically using an entropy maximization principle, 
as well as via a first-principles calculation using an  eigenfunction expansion in the continuum of low-energy states. A generalized  virial theorem
is derived, showing how the virial coefficient describes the delay in the diffusive spreading of the particles, found at large
distances. 
 When the process is non-recurrent, e.g. diffusion in three dimensions with a Coulomb-like potential, we use weighted time averages to restore basic canonical relations between time and ensemble
averages. 
\end{abstract}
\maketitle

\section{Introduction}
 
 The overdamped stochastic dynamics of a particle in an external potential field \eq{V(x)}, {in one dimension,} in contact with a thermal bath, is given by the Langevin equation
\begin{equation}
{\textrm{d} x \over {\textrm{d} t}} = - {V'(x) \over \gamma} + \sqrt{ 2 D} \eta(t).
\label{eq08}
\end{equation}
Here $-V'(x)$ is the deterministic force applied on the particle due to the potential field, and 
$\gamma>0$ is the friction constant. $\eta(t)$ is the bath noise, which is white, Gaussian,  has zero mean and
$\langle \eta(t)\eta(t') \rangle =\delta(t-t')$ (where $\delta(\cdot)$ is the Dirac \eq{\delta}-function).
The Einstein relation $D= k_B T /\gamma$ guarantees that the system, in the case
of a {\em{binding}} potential $V(x)$, will relax to thermal equilibrium.   
In this case, the steady-state equilibrium density is \cite{chandler1988introduction} 
\begin{equation}
P_{{\rm eq}}(x) = { \exp\left[ - V(x)/ k_B T\right] \over Z}.
\label{eq01}
\end{equation}
This final equilibrium state transcends a particular type of dynamics, and the asymptotic shape of the density does not depend on transport coefficients, such as the diffusion constant $D$, of the particles
in the medium. Here, 
\begin{equation}
Z= \int_{-\infty} ^\infty \exp\left[ - V(x)/ k_B T\right] {\rm d} x
\label{eq02}
\end{equation}
is the normalizing partition function, and $k_B$ is the Boltzmann constant.


A finite value of $Z$ is, however, not always guaranteed. In particular, $Z$ {can diverge when $V(x)$ generates a force $F(x)=-V'(x)\to 0$ when \eq{x\rightarrow\infty} and/or \eq{-\infty}. More specifically, in this manuscript we are interested in the case where \eq{V(x)} itself drops to zero at large distance, at least as fast as \eq{1/x}}. We initiated a study of this case in a previous work~\cite{aghion2019non}, finding that  at long times, and finite \eq{x}, the time-dependent density assumes the shape of the Boltzmann-Gibbs factor, multiplied by a factor which decays as power-law in time. In the limit \eq{t\rightarrow\infty}, the Boltzmann-Gibbs factor becomes an infinite-invariant density \cite{aghion2019non}. In the potential-free region, the density is simply the free-diffusion kernel. The appearance of the Boltzmann-Gibbs density can be understood as resulting from the fact that the particle returns infinitely many times to the potential region and so a kind of conditioned equilibrium is established there. {In addition, we then showed in \cite{aghion2019non} how in this thermal setting, we recover the Aaronson-Darling-Kac theorem \cite{darling1957occupation,aaronson1997introduction}, which describes the ergodic properties of a certain class of observables, integrable with respect to the infinite density. } In the current work,
we extend our previous study in several directions. Most notably, we re-derive our previous (one-dimensional) results using an eigenfunction expansion of the relevant time-dependent Fokker-Planck equation and thereby not only succeed in  recovering the infinite-invariant density, but the leading-order corrections as well. We also derive a virial theorem for our system, and extend some of our results to \eq{d\geq1} dimensions.

\begin{figure*}[t]
    \centering
    \includegraphics[width=1.0\linewidth]{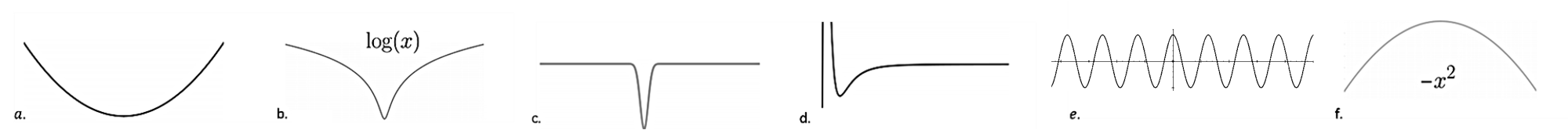}
    \caption{{\footnotesize{Various classes of potential landscapes: a. {An example of a binding potential; \eq{V(x)\sim x^2}. Here, the standard Boltzmann-Gibbs equilibrium state is achieved in the long-time limit, and standard ergodic theory applies.} b. A logarithmic potential, \eq{V(x) \sim \log (|x|)} for \eq{|x|\gg1}; here, the Boltzmann-Gibbs distribution is normalizable if the temperature is sufficiently small, otherwise it leads to non-normalizable Boltzmann-Gibbs statistics \cite{dechant2011solution,aghion2019non,bouin2020diffusion}. c.+d. Two-sided and one-sided asymptotically flat potentials always lead to a non-normalizable Boltzmann-Gibbs state, Eq. \eqref{EqID5}, and infinite-ergodic theory (see also \cite{aghion2019non,wang2019ergodic}). e. A periodic potential, with a structure that {stretches} for \eq{x\in(-\infty,\infty)} (also here one finds a non-normalised state, see e.g., \cite{sivan2018probability}). f. An unstable potential (see e.g., \cite{martin2018diffusion,Holubeck2019living}). All the potentials c-e, share the common property that in one dimension the process \eq{x(t)} in Eq. \eqref{eq08} is recurrent, however the mean return-time from large \eq{x} into any finite region around the origin is infinity. Thus, these potentials are weakly binding. This a sufficient, albeit not necessary condition for the emergence of an infinite-invariant density.}}}  
    \label{Fig1}
\end{figure*}

  {There are many other situations where \eq{Z} diverges as well. Some examples are presented in Fig. \ref{Fig1}, to be  compared with the  binding potential Fig. \ref{Fig1}a which leads to finite \eq{Z}. For logarithmic potentials, such as the example in Fig. \ref{Fig1}b, the partition function diverges when the depth of the well is sufficiently shallow. This happens when \eq{V(x)\sim V_0\log(|x|)} at large \eq{|x|}, and \eq{V_0<k_BT} at a certain given temperature, and the infinite invariant density  of this class of potentials was studied in  \cite{dechant2011solution,aghion2019non,bouin2020diffusion}). In addition, one may consider other non-binding fields, for example periodic structures, Fig. \ref{Fig1}e, random potentials, or unstable fields, Fig. \ref{Fig1}f. All the examples c-e share two properties: first, the partition function diverges, and second, the Langevin dynamics is recurrent in one dimension. In turn,  in one dimension 
this implies that the mean return time in these examples is diverging. 
We will refer to the class of potentials which fulfill these conditions as  \textit{weakly binding}. 
Logarithmic potentials are a marginal case, which behaves as binding   or weakly binding given the system parameters, and the unstable potential in Fig. \ref{Fig1}f belongs to neither group. As will become clear below, while  binding potentials lead to standard ergodic theory, we anticipate that infinite ergodic theory will serve as a useful tool for Langevin dynamics in one dimension for all the weakly binding  potentials (though as mentioned, in this work we study in detail only asymptotically flat fields). Note that the extension of the theory to higher dimensions is  not only a technical issue, since  in that case the random walk is no longer recurrent. The basic tools to deal with 
this case need some modifications, as we demonstrate for isotropic  potentials whose amplitude drops to zero at large radial distances \eq{r}, at least as fast as \eq{1/r}, below. 
} 


\begin{figure}[t]
\centering
\includegraphics[width=1.0\linewidth]{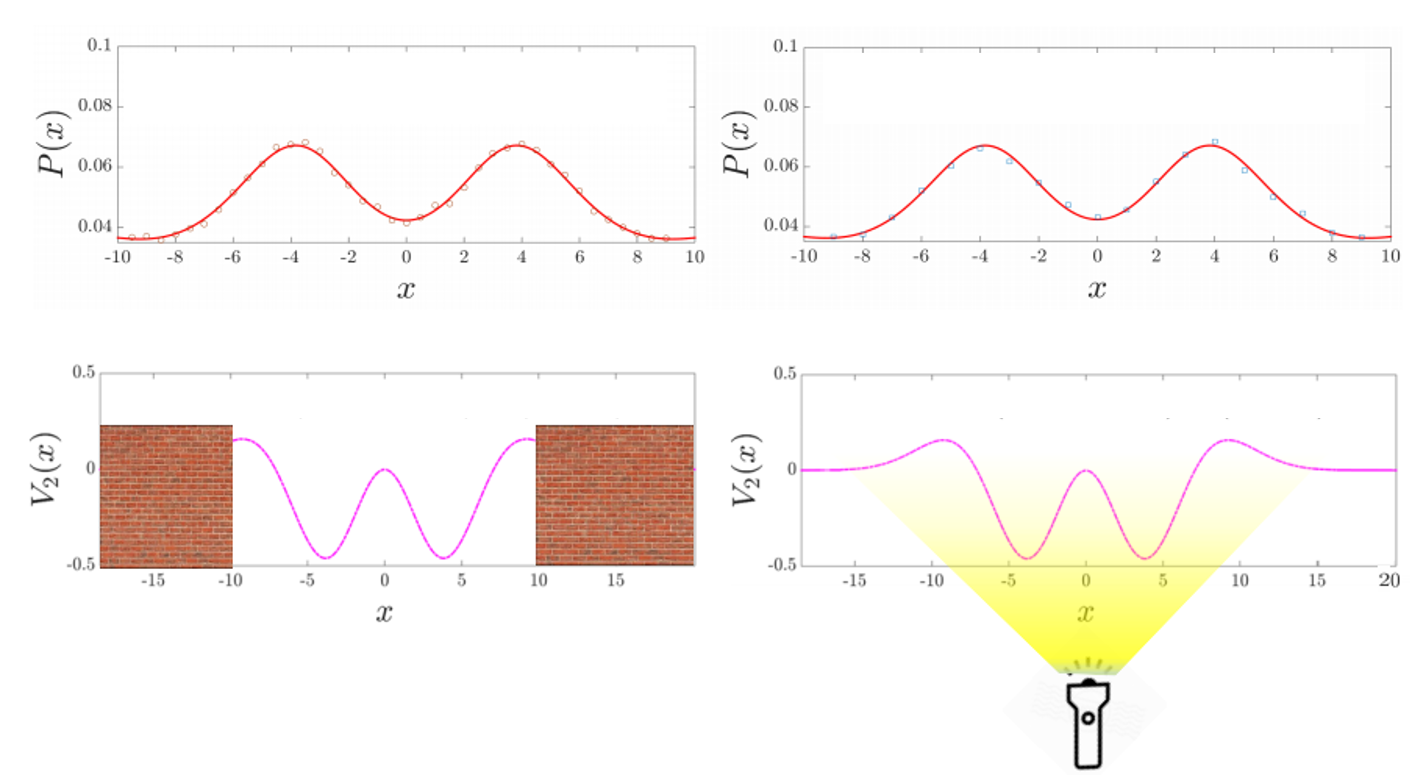}
\caption{{\footnotesize{In an open system, stretching from \eq{-\infty} to \eq{\infty}, the potential \eq{V(x)=[(x/5)^4-2(x/5)^2]\exp[-(x/5)^2]} is weakly binding, like the one seen in Fig. \ref{Fig1}c. There are two ways by which one can obtain the Boltzmann distribution in a system with such a potential.  Left bottom corner: we place the system within hard walls at \eq{\pm 10}, and since the system has a finite size, it has to eventually relax to equilibrium. Upper left corner: we see the simulated position distribution of the particles (blue circles), where the normalization is obtained by dividing the histogram of the positions by the number of particles and a bin-size. The theoretical red curve corresponds to Eq. \eqref{eq01}. Right bottom corner: when we remove the hard walls, particles escape freely to infinity. This case is very different from the previous one as here the system size is infinite. However, here we track only the section of \eq{-10<x<10}, and the particles which are found there at time \eq{t}. Upper right corner: The simulated position distribution of the particles, normalized only with respect to the particles that are found within the ``illuminated" region of \eq{x\in[-10,10]} (blue squares) at time \eq{t=10000}, versus the same theoretical red curve as above. 
As predicted by our previous results in Ref.~\cite{aghion2019non}, the upper two pictures are practically indistinguishable, despite describing very different physical scenarios. Our goal in this paper is to investigate in depth the dynamics of the latter scenario.}}}
\label{Fig2}
\end{figure}  

The structure of the manuscript is as follows: We first discuss in Sec. \ref{Prelim} some preliminary matters,  presenting a brief recap  of equilibrium statistical mechanics for binding potentials, where the partition function is normalizable, together
with a description of the particular examples of asymptotically flat potentials we use for our simulations. 
In Sec. \ref{NnBGState} we define the non-normalizable Boltzmann state. In Sec. \ref{SecEntropyEnergy}, we discuss the entropy maximization principle. In Sec. \ref{SubSecEigen}, we provide the derivation of the leading-order time-dependent shape of the particle density and obtain higher-order correction terms.  In Sec. \ref{SecTimeAndEnsembleMeans}, we discuss time and ensemble averages, and in Sec. \ref{FluctuationsOfTimeAverages} we discuss the fluctuations of the time average and infinite ergodic theory. In Sec. \ref{VirialTheorem125}, we study the virial theorem{. In Sec. \ref{dDimenssional}, we show that the non-normalizable Boltzmann state exists in any dimension \eq{d\geq1}, and extend our analysis of the ratio between time and ensemble averages of integrable observables, to any dimension. {A summary of our main results is found in Sec. \ref{SectionReview}.} The} discussion is found in Sec. \ref{Discussion}.   

\section{Preliminaries\label{Prelim}}
\subsection{A recap of statistical mechanics} 
Before treating weakly binding potentials,
we first recall the standard treatment of the case where $V(x)$ is increasing with distance in such a way
that the partition function is finite, e.g., $V(x)=x^2$ (a binding potential, Fig.~\ref{Fig1}.a). 
According to the basic laws of statistical
physics, the system is ergodic (we assume that $V(x)$ does not divide the system into
compartments). Let ${\cal O}[x(t)]$ be a physical observable. Then, in the long-time limit,
\begin{equation}
\overline{{\cal O}[x(t)]}  \rightarrow
\langle {\cal O}(x) \rangle.  
\label{eq03}
\end{equation}
Here, the overline denotes a time-average $\overline{{\cal O}[x(t)]} = \frac{1}{t}{\int_0 ^t {\cal O}[x(t')] {\rm d} t'}$, and the brackets an ensemble-average. 
The fact that the time-average, which is what is measured
in many experiments, converges to the corresponding ensemble-average (in the long-time limit), is very useful
for the theoretician, who usually considers the latter;
\begin{equation}
\langle {\cal O}(x) \rangle = { \int_{0} ^\infty {\cal O}(x) \exp[ - V(x)/k_B T ] {\rm d} x \over Z}.
\label{eq04}
\end{equation}
{It should be noted that some observables, such as  
$\mathcal{O}(x) = \exp[V(x)/T]$, are not integrable with respect to the Boltzmann distribution,
however these are mostly not the main focus of physicists.} 

 Statistical mechanics is related to thermodynamics in many textbooks.
In particular,   the Helmholtz  free energy is
\begin{equation}
{\cal F}  = \langle V \rangle - T S,
\label{eq05}
\end{equation}
where we omitted the thermal kinetic energy term without any loss of generality.
The entropy is 
\begin{equation}
S(t) = - k_B \int_0 ^\infty P_t(x) \ln [ P_t(x)] {\rm d} x  ,
\label{eq051}
\end{equation}
where $P_t(x){\rm d} x$ is the probability of finding the particle at time $t$ in the interval 
$(x,x+{\rm d} x)$. In equilibrium, we take the long-time limit, and for generic initial conditions
we have
\begin{equation}
\lim_{t \to \infty} P_t(x) = P_{{\rm eq}} (x) ,
\label{eq06}
\end{equation}
so, in this limit,
\begin{equation}
{\cal F} = - k_B T \ln Z, 
\label{q07}
\end{equation}
and 
\begin{equation}
P_{{\rm eq}} (x) = { e^{ - V(x)/k_b T}\over e^{- {\cal F}/k_b T}}.
\label{eq07}
\end{equation}
The  Boltzmann factor appearing in the numerator is the essence of the canonical
ensemble, while in the denominator we have the relation to thermodynamics. The goal of this manuscript is to consider the  case where $S$ is increasing with time,
as opposed to saturating to a limit, but still all this structure remains intact when the appropriate modifications
are made, namely 
we must use the tool of  non-normalizable Boltzmann-Gibbs statistics \cite{dechant2011solution,aghion2019non,wang2019ergodic}. This idea, which is discussed at length below, harnesses the tools of infinite-ergodic theory, which has been well established as a mathematical theory for several decades, and {was discovered} in recent years also in physical systems, e.g., \cite{darling1957occupation,aaronson1997introduction,zweimuller2009surrey,thaler2001infinite,akimoto2010role,korabel2012infinite,klages2013weak,akimoto2015distributional,meyer2017infinite,leibovich2019infinite,akimoto2019infinite,zhou2019continuous,sato2019anomalous,radice2020statistics}. {Note that infinite, namely non-normalizable, densities serve two main goals; the first is for computation of large deviations and rare-event statistics of fat-tailed stochastic systems, and the second is in the context of ergodic theory. In that regard, one should distinguish between infinite invariant and infinite covariant densities, 
the latter are not discussed here (see e.g., Refs. \cite{kessler2010infinite,lutz2013beyond,rebenshtok2014non,rebenshtok2014infinite,holz2015infinite,aghion2017large,wang2019transport,vezzani2019single}).}

\subsection{Asymptotically flat potentials}
 
As mentioned above, {in this work we treat the class of potentials where  \eq{V(x)\rightarrow0} at \eq{x\rightarrow\infty}, and the drop rate is at least as fast as  \eq{1/x}.  In this case \eq{V(x)} belongs to the larger class of potentials which are weakly binding. Our leading-order results below apply to any such asymptotically flat potential; however, some of our calculations of higher-order correction terms below are obtained only for potentials that fall-off faster than \eq{1/x^2}.} Furthermore, we distinguish between two situations: the one-sided case, where $V(x)=\infty$ when $x\leq0$, and the double-sided system, where, to simplify explanations, we assume that the system is symmetric, and so  \eq{V(x)\rightarrow0} also when \eq{x\rightarrow-\infty} (but our results can be  trivially extended also to the non-symmetric case). The first situation can be realized experimentally, for example, when the particles are diffusing in three dimensions in a heat bath, above a flat surface, and their interaction with that surface is given by \eq{V(x)}, where \eq{x} is their height (see e.g., \cite{pavani2009three,chechkin2012bulk,metzler2014anomalous,campagnola2015superdiffusive,krapf2016strange,wang2017three}). Here, the potential is infinite when \eq{x} is zero, since the surface is impenetrable. The second case, corresponds e.g., to a scenario where the particles diffuse in a liquid while being loosely held by optical tweezers. The intensity of such an optical trap drops with the distance from the potential well, often like an inverted Gaussian \cite{ashkin1986observation, grier2006holographic,DrobczynskiOpticalTweezers}. 

In our simulations,  we mostly used the following two examples: the one-sided Lennard-Jones type potential (which depends only on the height coordinate, $x$, in the scenario of three dimensional diffusion above a hard-surface):  
\begin{equation}
V(x) = 
\left\{ 
\begin{array}{c c}
\infty & x\leq0 \\
V_0 \left[ \left( { a \over x}\right)^{12} - \left({ b \over x}\right)^6\right]
& x>0
\end{array}
\right. ,
\label{eq09}
\end{equation}
where $V_0$, $a$ and $b$ are positive constants, and the symmetric potential (which can be realized using optical tweezers e.g., \cite{grier2006holographic,DrobczynskiOpticalTweezers}):
\begin{equation}
V(x) = 
\left[(x/5)^4-(x/5)^2\right]e^{-(x/5)^2}.
\label{eq091}
\end{equation}
In both cases, there is no thermal equilibrium in the usual sense. Famously, this ``problem" with asymptotically flat potentials was  pointed out by Fermi already in $1924$ \cite{fermi1924wahrscheinlichkeit}. Physically, when $x$ is large, the deterministic force is negligible and then the particles are diffusing in the bulk.
Our discussion below is not limited to a specific form of the potential field, provided that it is eventually flat, but the key assumption is that the fluctuation-dissipation relation applies, namely the Einstein relation is valid. This fact, as we show, allows for modified thermal concepts to emerge, even though $Z=\infty$.  
Note that unless specified explicitly, below we present our derivations mostly for the one-sided potentials, just for simplicity of writing. 

\subsection{Limitations of the standard treatment of the non-normalizability of the partition function, and the alternative}

The standard response to the ``problem" of the divergence of the Boltzmann factor, for any type of potential, is to introduce a finite size to the system, $L$, and since then the limits
of the integral in Eq. \eqref{eq02} will
stretch only up to this limit, it is guaranteed to have a finite value. In this case, the system will {always} approach equilibrium in the long-time limit. On the bottom-left panel of Fig. \ref{Fig2}, we see an example of a system enclosed between two walls, with a potential that, without the walls (namely, if the system had stretched from \eq{-\infty} to \eq{\infty}), would have been weakly binding. On the top-left panel of the figure, we see the corresponding Boltzmann distribution of the particles. But in this work, we do not wish to impose this constraint. {One reason is simply that many experimental settings do not have truly hard walls or confining potentials 
e.g., when using optical traps  (see for example, \cite{grier2006holographic,DrobczynskiOpticalTweezers}). The second reason is that even when the systems is confined by its boundaries, very often particle tracking experiments are not long enough for the particles to encounter them.} 

The results of our previous work, Ref.~\cite{aghion2019non}, imply that we can take a very different physical approach, but still retain the exact same shape of $P_t(x)$ at long times, obtained in the hard-wall scenario.
In this second approach, we focus our range of observation to a single slice of space, and normalize the probability distribution only with respect to the number of particles that are present in this region at time \eq{t} (namely, by discarding the particles which are found outside of this region). Note, that 
this situation is common in single-particle-tracking experiments, where the microscope in use has a finite field of view, hence this approach is essentially similar to ``looking under the lamp". The result of a simulation of this is shown in the right side of Fig. \ref{Fig2}, to be compared with the wall scenario on the left. The measured concentration of the remaining particles is identical (up to statistical fluctuations) to that of the equilibrium state of the ``walled" system, even though the distributions come from two different physical setups. As we shall see in the following, the dynamics by which the two final distributions are attained are extremely different. The ``walled" system converges exponentially in time (with the time diverging as $L\to\infty$), whereas the alternative converges as a power-law.

\section{The non-normalized Boltzmann-Gibbs state}
\label{NnBGState}
In this section, we review the derivation of the long-time limit of the distribution function to leading-order. We then analyze its thermodynamic implications. We start from the Fokker-Planck equation description of the diffusion process controlled by the Langevin Eq. (\ref{eq08}), which specifies the dynamics of 
the  concentration
of particles, or equivalently the probability density function $P_t(x)$, 
\begin{equation}
{\partial P_t(x) \over \partial t} = D \left[ {\partial^2 \over \partial x^2} + {\partial \over \partial x} { V'(x) \over k_B T}\right] P_t(x).
\label{eq10}
\end{equation}
We treat this equation in the long-time limit. If we set the left-hand side to zero, namely we search for 
a time-independent solution, which we call ${\cal I}(x)$,  we have
\begin{equation}
0 =  \left[ {\partial^2 \over \partial x^2} + {\partial \over \partial x} { V'(x) \over k_B }T\right] {\cal I} (x),
\label{eq11}
\end{equation}
and hence one appealing  option reads
\begin{equation}
{\cal I}(x) = \mbox{Const} \exp\left[- V(x)/k_B T\right].
\label{eq12}
\end{equation}
However, this solution does not satisfy the boundary condition $\lim_{x\to \infty} P_t(x)=0$
(unless $\mbox{Const}=0$)
and is clearly non-normalizable when \eq{V(x)} is asymptotically flat at large distances. This is certainly not a possibility as the particles
are neither created nor annihilated, so the normalization is conserved
$\int_0 ^\infty P_t(x) {\rm d} x = 1$ for any $t\ge 0$. In-fact, mathematically as we will show below, the non-normalized solution
${\cal I}(x)$ is an infinite-invariant density \cite{aaronson1997introduction}, as opposed to a probability density. 

\begin{figure}[t]
\includegraphics[width=1.0\linewidth]{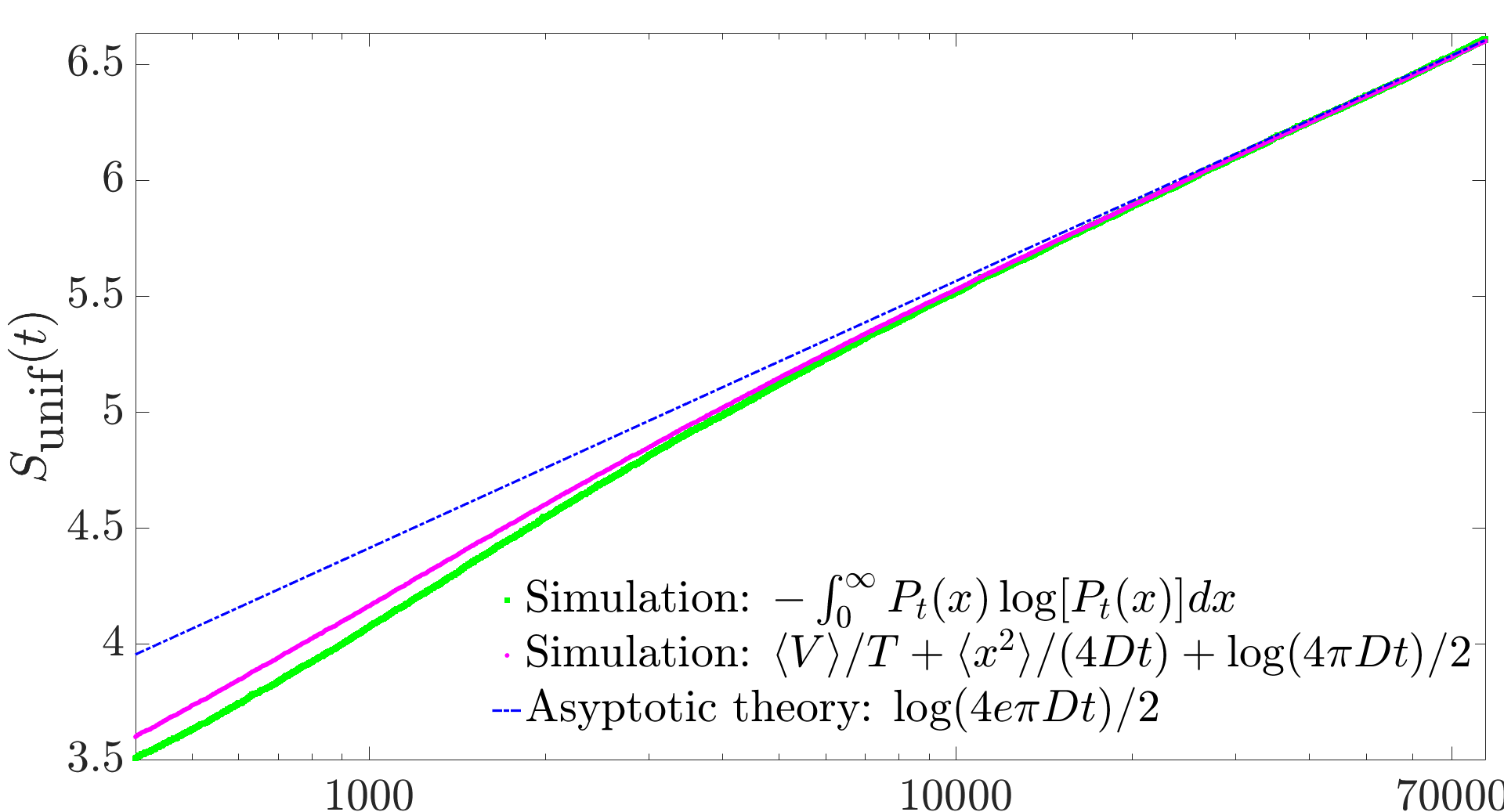}
\caption{{\footnotesize{ (color online) The Gibbs entropy,~$S(t)=-\int_0^\infty P_t(x)\log{[P_t(x)]}\Intd x$, obtained from simulation results of the overdamped Langevin Eq. \eqref{eq08}, and the symmetric potential Eq. \eqref{eq091}, for $10^5$ particles, at $k_BT=0.2$ (green squares). The purple circles corresponds to Eq. \eqref{EntropyEnergy} (but with \eq{\pi\rightarrow4\pi}, since the potential is symmetric), where all the observables are measure from the simulation. Here, one should note that additional correction terms of order \eq{1/\sqrt{t}} might exist, due to contribution from correction terms to the leading-order behavior of \eq{P_t(x)}, discussed in Sec. \ref{SubSecEigen}. Such correction terms will decay in time as fast as \eq{\langle V\rangle}. The blue line corresponds to the asymptotic theory, Eq.~\eqref{eq16} (with \eq{\pi\rightarrow4\pi}), which becomes exact when $t\rightarrow\infty$.}}}
\label{SM22}
\end{figure}

 Since \eq{\mathcal{I}(x)} is non-normalized, we search for a more complete solution in the form of 
\begin{equation}
P_t(x) \sim { A e^{ - {V(x) \over k_B T}} \over t^\alpha}
\label{eq13}
\end{equation}
with $\alpha>0$. {The logic behind this ansatz is that, instead of Eq. \eqref{eq12} which is obtained from Eq. \eqref{eq11} where the left-hand side is exactly zero, we now look for a time-dependent solution to Eq. \eqref{eq10}, where the contribution from the left-hand side is non-zero, yet small.} This long-time behavior, when inserted into the Fokker-Planck equation,
is a solution to leading-order, with correction terms
of order $\partial_t t^{-  \alpha} \propto t^{-(1+\alpha)}$ which are smaller by a factor of $1/t$ than the
leading $t^{-\alpha}$ term. Physically, we can expect this solution to be valid
only for $x\ll l(t)$, where $l(t) = \sqrt{ 4 D t}$ is the diffusion length-scale of the problem.
In the  range $x\gg 1$, on the other hand, we know that the force is negligible.
Hence, in that case,
\begin{equation}
P_t(x) \simeq {1 \over \sqrt{\pi D t}} \exp[ - {x^2 \over 4 D t}].
\label{eq14}
\end{equation}
Matching Eqs. (\ref{eq13},\ref{eq14}) in the overlap region $1 \ll x \ll l(t)$, we find
$\alpha = 1/2$ and $A = 1/\sqrt{ \pi D}$. 
A uniform approximation then reads
\begin{equation}
P_t(x) \simeq { 1 \over \sqrt{ \pi D t}} e^{ -  {V(x) \over  k_B T}} e^{ -  {x^2 \over 4 D t}},
\label{UniformApproximation}
\end{equation}
where we set $V(\infty)=0$.
 This scaling solution is valid at long times for all $x$. For a process with a potential of the form in Eq.  \eqref{eq091}, where the particle is allowed to cross also to \eq{x<0}, the factor \eq{\sqrt{\pi D t}\rightarrow 2\sqrt{\pi D t}}, and similarly \eq{A\rightarrow A/2} in Eq. \eqref{eq13}. In Sec. \ref{SubSecEigen} we derive this solution for any potential which decays faster than \eq{1/x^2} at large distances, using an eigenfunction expansion method that employs the continuous spectrum of the Fokker-Planck operator.  This method also yields correction terms that vanish faster than \eq{1/\sqrt{t}}. We leave the equivalent derivation for the case of potentials that fall off like \eq{1/|x|^\zeta}, {where \eq{1<\zeta\leq2}}, out of this manuscript, since here all the results associated with the leading order behavior in time are similar to the \eq{\zeta>2} case, but the correction terms are different. 

Importantly, to obtain Eq. \eqref{UniformApproximation}, we have assumed that the particles are initially
localized, say on  $x_0$ or within an interval $(0,x_0)$, or more correctly the
initial density has at least an exponential 
cutoff. The scale $x_0$ does not alter the long-time solution (to leading-order approximation, see Sec. \ref{SubSecEigen}), and  since the solution forgets its initial conditions,
we introduce a thermodynamic notation
\begin{equation}
P_t(x) \propto   e^{ -  \beta V(x) - \xi  x^2 }
\label{eqnew}
\end{equation}
with $\beta=1/k_B T$ and $\xi=1/4 D t$. 

\begin{figure}[t]
\centering
\includegraphics[width=1.0\linewidth]{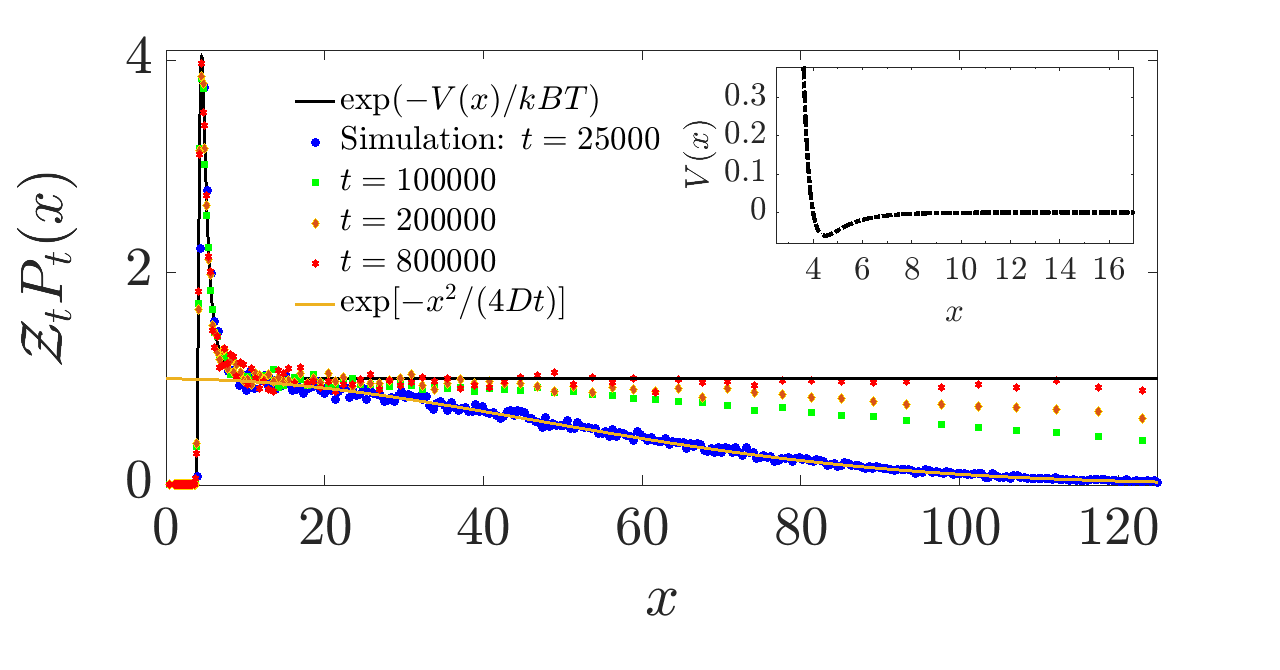}
\includegraphics[width=1.0\linewidth]{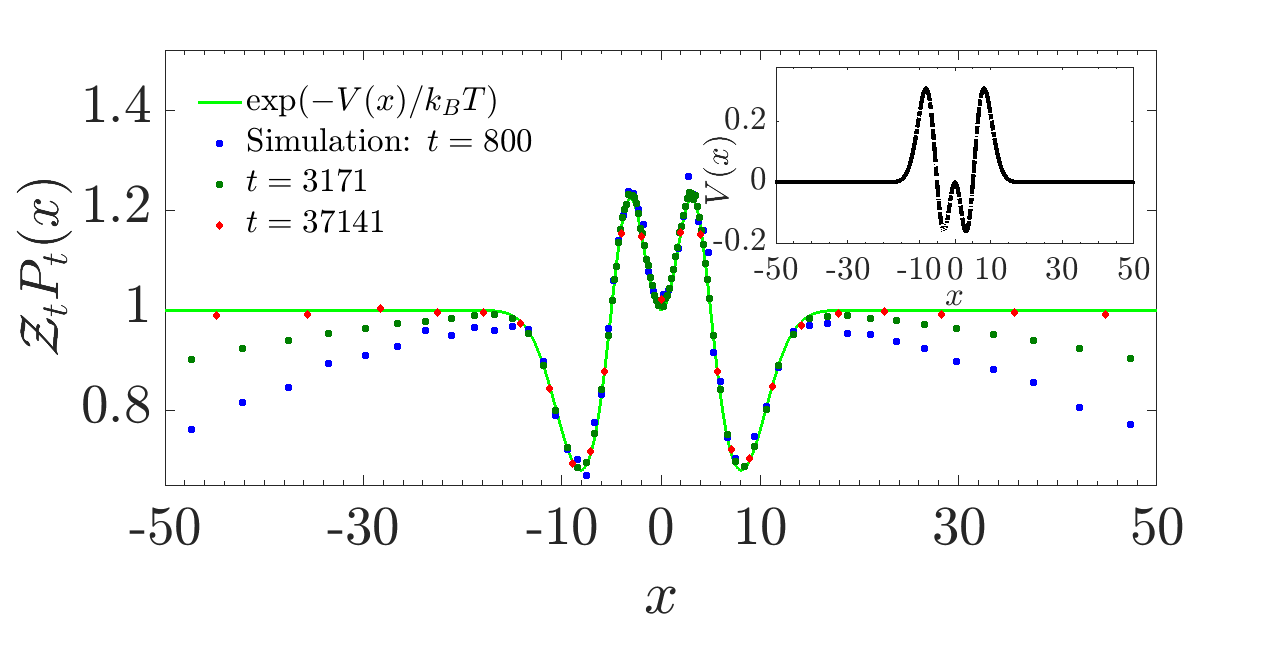} 
\caption{{\footnotesize{(Color online) \textit{Upper panel:} The non-normalizable Boltzmann state Eq. \eqref{EqID5}, corresponding to the Lennard-Jones potential, Eq. \eqref{eq09}; Theory in black line, simulation results for \eq{\mathcal{Z}_tP_t(x)}, at various increasing times \eq{t}, where \eq{\mathcal{Z}_t=\sqrt{\pi D t}} (\eq{k_BT=~0.0436}, \eq{\gamma=1}), appear in colored symbols. The inset shows the potential, where \eq{U_0=1000} and \eq{(a,b)=(1,2)}. At finite times one can see deviations from the asymptotic Boltzmann state, which vanish as \eq{t} increases. As seen by the yellow line (at e.g., \eq{t=2500}), at large \eq{x}, \eq{P_t(x)} is Gaussian (since the attractive force is zero). \textit{Lower panel:} The non-normalizable Boltzmann state corresponding to the two-sided potential Eq. \eqref{eq091}, with \eq{(k_BT,\gamma)=(0.8,1)} (green curve). Here \eq{\mathcal{Z}_t\rightarrow{2\mathcal{Z}_t}}. Simulation results in colored symbols  (and the inset shows the potential).}}}
\label{SM25}
\end{figure}

 We now consider the entropy of the system.  
Using the uniform approximation, Eq. (\ref{UniformApproximation}), we find 
\begin{equation}
S_{\mbox{unif}}(t) =  k_B \int_0 ^\infty {\rm d} x P_t(x) \left[ \ln \sqrt{\pi D t} +  \beta  V(x)  + \xi x^2 \right].
\label{eq16a}
\end{equation}
This yields 
\begin{equation}
S_{\mbox{unif}}(t) = {\langle V \rangle \over T}  + k_B \xi \langle x^2 \rangle - {\cal {\tilde F}}/T, 
\label{EntropyEnergy}
\end{equation}
where the averages are with respect to Eq. \eqref{UniformApproximation}. 
The average energy $\langle V \rangle$ and $\beta$ 
are thermodynamic pairs, and at least suggestively the same is true
for $\langle x^2 \rangle$ and $\xi$. Later, we will make this analogy
to equilibrium statistical mechanics more precise, using an entropy maximization formalism. Here, the free energy is ${\cal \tilde{F}} = - (k_B T/2) \ln ( \pi / 4 \xi)$.
It stems from the first term in Eq. (\ref{eq16a}), which in the usual setting
gives the connection between the Helmholtz free energy and the 
partition function. We now see that, for a fixed observation time and \eq{D}; Eq. \eqref{EntropyEnergy} means that {$1/T = (\partial S_{\mbox{unif}}/\partial \langle V\rangle)$}, in agreement with {standard} thermodynamics. 
To leading order, using the fact 
 that the mean-squared displacement is behaving diffusively at long times, namely 
$k_B \langle x^2 \rangle / 4 D t=  k_B /2$, we have  
\begin{equation}
S_{\mbox{unif}}(t) \sim {k_B \over 2} - {\cal {\tilde F}}/T={k_B \over 2} \ln (\pi e  D t) .
\label{eq16}
\end{equation}
In this limit, the entropy is insensitive to the potential, since the average potential energy approaches zero at increasing times. This occurs
simply because the particles increasingly explore the large \eq{x} region where the potential is flat (zero). Below, we study the average potential energy in detail, but for now it is only
important to realize that entropy times $T$ is far larger. 
Eq. (\ref{eq16}) shows that 
the entropy is increasing with time, which is to be expected, 
 since the packet of particles is spreading out to the
medium. Fig. \ref{SM22} shows the entropy versus time, obtained from simulation results using the two-sided potential, Eq. \eqref{eq091} (green squares), and the corresponding measurement based on Eq. \eqref{EntropyEnergy} (purple circles). But in the latter, since the potential is symmetric, \eq{S_{\mbox{unif}}(t)\sim \frac{k_B}{2}\ln{(4e\pi Dt)}} (the reason is that at the tails, the density is now proportional to \eq{\exp[-x^2/(4Dt)]/\sqrt{4\pi Dt}}). It also shows the asymptotic logarithmic growth at long times (blue line), based on Eq. \eqref{eq16}, but with \eq{\pi\rightarrow4\pi}. {The figure shows that Eq. \eqref{EntropyEnergy} is a good approximation, but note that additional correction terms of order \eq{1/\sqrt{t}} might exist, due to contribution from correction terms to the leading-order behavior of \eq{P_t(x)}, discussed in Sec. \ref{SubSecEigen}, which will decay in time as fast as \eq{\langle V\rangle}. Even in this case, the asymptotic logarithmic growth at long times, seen in Eq. \eqref{eq16}, will remain unchanged.}

We are now ready to define the non-normalized Boltzmann density more precisely.
We define the time dependent function
\begin{equation}
\mathcal{Z}_t = \exp(- {\cal \tilde{F}}/k_B T),
\label{eq19}
\end{equation}
hence in our case $\mathcal{Z}_t = \sqrt{ \pi D t}$ (or \eq{\sqrt{4\pi Dt}} for two-sided potentials). 
Inserting \eq{\mathcal{Z}_t}  into Eq. 
(\ref{UniformApproximation}),
we find
\begin{equation}
\lim_{t \to \infty} \mathcal{Z}_t P_t(x) = \exp[ - V(x)/k_B T].
\label{EqID5}
\end{equation}
Here we used $\lim_{t \to \infty} \exp[- x^2/4 D t]=1$ for any finite $x$,
though in a finite-time experiment, this identity will be valid for $x\ll l(t)=\sqrt{4Dt}$. 

At least in principle, with  a measurement of the entropy in the long-time limit,  
which is possible with an ensemble of particles, and using Eqs. (\ref{eq16},\ref{eq19}), 
we can determine $\mathcal{Z}_t$. And with that information, we can verify Eq. (\ref{EqID5}).
Note that, clearly, if we know $D$ up front, we can find $\mathcal{Z}_t$
without measuring entropy at all,  but one point of our work is to claim that the thermodynamic 
formalism may have a more general validity, which is a question worthy of further study. 
Since $\mathcal{Z}_t$ is increasing with time, then when it is multiplied by 
the normalized density $P_t(x)$, this leads to a non-normalizable Boltzmann factor, when $t$ becomes large. 
Again, mathematically, the non-normalized Boltzmann factor is the infinite-invariant density  of the system \cite{aaronson1997introduction,aghion2019non}. Of course Eq. (\ref{EqID5}) holds also for the case when $\mathcal{Z}_t$ eventually
approaches a constant, namely for finite size systems. {Fig. \ref{SM25}, shows simulation results for  \eq{\mathcal{Z}_tP_t(x)}, corresponding to the potentials in Eq. \eqref{eq09} (top figure), and Eq. \eqref{eq091} (bottom). At increasing times, the scaled distributions approach  the non-normalizable Boltzmann state, Eq. \eqref{EqID5}, while the finite-time Gaussian tails where \eq{x\sim\sqrt{t}}, are pushed increasingly outward. The insets show the potentials.} 
Note that we obtain the correction terms to the leading behavior of \eq{P_t(x)} in Sec. \ref{SubSecEigen}, but for now we leave the rest of the discussion about the corrections to the mean energy and entropy out of the scope of this manuscript. The benefit of that discussion is in the rigorous derivation of the shape of \eq{P_t(x)} and the effect of the initial condition, which is shown to be negligible (namely it does not alter Eq. \eqref{EqID5}). 

Using Eq. (\ref{EqID5}), one can determine the shape of the potential field in the system from the position density of the particles. We do not, however, address the question of whether this yields
 the mechanical or electrostatic force,
or an effective force, as clearly the potential of the mean force might itself be temperature-dependent.  


\section{Entropy extremum principle}
\label{SecEntropyEnergy} 

The structure of the theory suggests that a more general principle is 
at work. The entropy extremum principle is a natural choice, 
with the imposition of three added constraints. These are: the normalization condition, a finite mean energy condition (this allows us to treat fluctuations
of energy in the canonical ensemble, unlike the microcanonical ensemble,
where the energy surface is fixed), and the special feature of 
our system, which is that
the mean-squared displacement is diffusive, $\langle x^2\rangle \sim 2 D t$. This latter constraint
is the new ingredient of the theory and this introduces the time dependency.  


 We define a functional of the density $P_t(x)$  at some fixed
large $t$ 
\begin{align} {\cal S} [P_t(x)] =& - k_B \int_0 ^\infty P_t(x) \ln P_t(x) {\rm d} x \nonumber\\ 
& - \beta k_B \left(\int_0 ^\infty V(x) P_t(x) {\rm d} x- \langle V\rangle\right) \nonumber\\ 
&
- \lambda k_B \left( \int_0 ^\infty P_t(x) {\rm d} x - 1 \right)  \nonumber\\ 
&
- \xi k_b \left( \int_0 ^\infty x^2 P_t(x) {\rm d} x - 2 D t\right).
\label{CON01}
\end{align}
Here  
$\beta,\lambda$ and $\xi$
are  Lagrange multipliers. 
The first term is the  entropy, which in the absence of the constraints
implies that all micro-states are  equally probable. If we set
$\xi=0$, and find the extremum, we get the usual Boltzmann-Gibbs theory,
however that can be valid only if the potential is binding, which is
not the case under study here.  Taking the functional derivative, we get
\begin{equation}
P_t(x) = N \exp\left( - \beta V(x) - \xi x^2\right),
\label{CON02}
\end{equation}
where $N$ is the normalization constant. The constraints are
\begin{align}
&N \int_0 ^\infty \exp\left[ - \beta V(x) - \xi x^2\right] {\rm d} x =1,\nonumber\\ 
&N\int_0 ^\infty x^2 \exp\left[ - \beta V(x) - \xi x^2 \right]  {\rm d} x = 2 D t, \quad\mbox{and}\nonumber\\ 
&  N \int_0 ^\infty V(x) \exp\left[ -\beta V(x) - \xi x^2\right] {\rm d} x = \langle V(x) \rangle.
\label{CON03}
\end{align}
These conditions are satisfied if $\xi$ is small and $V(x)$ is asymptotically
flat. To see this, we use $\xi  x^2= y^2$, so we can rewrite the
first two integrals  as:
\begin{equation}
{N \over \sqrt{\xi} }  \int_0 ^\infty  \exp\left[ - \beta V(y/\sqrt{\xi}) -
y^2\right] {\rm d} y =1
\label{eqBLA}
\end{equation}
and
\begin{equation}
{N \over \sqrt{\xi}^3 }  \int_0 ^\infty y^2  \exp\left[ - \beta V(y/\sqrt{\xi}) -
y^2 \right] {\rm d} y  = 2 D t, 
\label{eqBLA1}
\end{equation}
and since $\lim_{\xi \to 0} V(y/\sqrt{\xi})=0$ we may ignore the
potential field in th{is} limit. Solving the Gaussian
integrals, we find
\begin{equation}
\xi = {1 \over 4 D t} , \ \ \ \  N = {1 \over \sqrt{ \pi D t}}.
\end{equation}
Notice that in Eqs. (\ref{eqBLA},\ref{eqBLA1}), we are averaging over
observables which are non-integrable with respect to the Boltzmann infinite
density, i.e. $x^0,x^2$ (or $y^0,y^2$). 
For the last constraint we use the small parameter $\xi$ to 
approximate
$ V(x) \exp[ - \beta V(x) - \xi  x^2] \to V(x) \exp[ - \beta V(x)]$, and 
since in any reasonable
 range where the potential is finite, $\exp[ - \xi x^2]$
is equal to $1$, we get $\langle V\rangle = N \int_0 ^\infty V(x) \exp[  - \beta V(x)] {\rm d} x.$
To summarize, we find that the extremum principle gives
\begin{equation}
P_t(x) \simeq {\exp\left[ - \beta V(x) - x^2 / 4 D t\right]
\over \sqrt{ \pi D t} }.
\label{MOOS}
\end{equation} 
This is the same as the uniform approximation Eq. 
(\ref{UniformApproximation}), which was proven valid for a specific stochastic
model, i.e. the overdamped Langevin equation. However, the 
extremum principle suggests that 
the existence of the non-normalized Boltzmann density is of more general validity, even outside of this particular context. 
Finally, one could claim that since thermodynamics is a theory which does not
depend explicitly on time $t$, we cannot identify $\beta$ with the inverse of the temperature. However, at least within our Langevin model, the Einstein 
relation and our analytical results
give both  the physical and the mathematical motivation to make this relation. 

\section{Eigenfunction expansion, and corrections to the uniform approximation}
\label{SubSecEigen}
We have obtained the leading-order behavior of \eq{P_t(x)}
at long times for  asymptotically flat potentials from two different directions, first by using physically inspired guesses for the small and large $x$ regimes and then matching, and secondly via entropy maximization. 
Here, we re-derive our result a third time, but importantly, now we use an eigenfunction expansion, so as to allow access to the leading-order corrections. In particular, we focus on potentials that fall off faster than \eq{1/x^2} at large \eq{x}.  The spectrum of the Fokker-Planck operator is continuous since the system is diffusive in the bulk. For convenience, we consider the case that there is a reflecting wall at $x=0$, giving rise to a no-flux boundary condition, $P_t'(0)=0$. The final answer works as well for the case that the potential diverges to $+\infty$ as $x\to 0$ from above, so that again the particles are restricted to $x>0$.
For a \eq{\delta}-function initial distribution, centered around some positive \eq{x_0}, we show that the initial condition does not affect the asymptotic shape to leading order in time, and we obtain the correction to the leading-order term where it first makes its appearance. 
Note that the eigenfunction expansion in the case of a two-sided potential follows along the same lines, but given the details of the setup one may need to examine both symmetric and non-symmetric solutions for the eigenmodes.  

Starting from the Fokker-Planck equation,
\EQ{\partial_t P_t(x)=\hat{L}_{\textit{FP}}P_t(x),}{FokkerPlanckEigenfunction} 
where \eq{\hat{L}_{\textit{FP}}=D[\partial^2_x+\partial_x U'(x)]} is the Fokker-Planck operator and \eq{U(x)=\frac{V(x)}{k_BT}}, using the unitary transformation~\cite{dechant2011solution} \eq{\hat{\mathcal{H}}=e^{U(x)/2}\hat{L}_{FP}e^{-U(x)/2}}, $\Psi(x,t)=e^{U(x)/2}P_t(x)$,  we write the following Schr\"odinger-like equation 
\EQ{\hat{\mathcal{H}}\Psi(x)=-\lambda\Psi(x).}{SMA1}
For technical reasons, we imagine an infinite effective-potential wall at large $x=L$, so that the spectrum is discrete. As the eigenvalues $\lambda$ are positive definite ($\lambda=0$ is not in the spectrum as it is not normalizable in the $L\to \infty$ limit, and hence the system does not reach equilibrium~\cite{dechant2011solution}), we write $\lambda\equiv Dk^2$, for some discrete set of $k$'s. We will eventually take the $L\to\infty$ limit, before we take the large $t$ limit.
Then,  
\EQ{P_t(x)=e^{-U(x)/2+U(x_0)/2}\sum_{\{k\}} N_k^2 \Psi_k(x_0)\Psi_k(x)e^{-Dk^2 t}.}{SMA2} 
It will prove convenient to set the normalization of $\Psi_k(x)$ via the condition  $\Psi_k(0)=e^{-U(0)/2}$, so we have to incorporate an explicit normalization factor $N_k^2$.
 Thus, Eq. \eqref{SMA1} translates to \small\EQ{\frac{\partial^2}{\partial x^2}\Psi_k(x)+\frac{U''(x)}{2}\Psi_k(x)-\frac{U'^2(x)}{4}\Psi_k(x)=-k^2\Psi_k(x).}{SMA3} \normalsize 
 with boundary conditions
 \begin{equation}
     \Psi_k(0)=e^{-U(0)/2}, \qquad \Psi_k'(0)=-\frac{U'(0)}{2}e^{-U(0)/2}.
     \label{eqBC}
 \end{equation}
    
The long-time limit is clearly dominated by the small-$k$ modes, so we need to consider only them. We need to treat two regimes separately,  first the range \eq{x\ll1/{k}}, where the right-hand side of Eq. \eqref{SMA3} is always small, (denoted region \textbf{I}), and second, for \eq{x\gg 1} 
(region \textbf{III}). We match the two asymptotic limits in the overlap region \eq{1\ll x\ll1/k}  (region \textbf{II}). 

In Region \textbf{I}, the term $- k^2\Psi_k(x)$ is negligible due to the smallness of $k$. To leading order, then, we have the homogeneous equation,
\small\EQ{\frac{\partial^2}{\partial x^2}\Psi_k(x)+\frac{U''(x)}{2}\Psi_k(x)-\frac{U'^2(x)}{4}\Psi_k(x)=0,}{SMA3a} \normalsize 
with the solution (satisfying the no-flux boundary conditions Eq. \eqref{eqBC} at $x=0$)
  \eq{\Psi_h(x)= e^{-U(x)/2},}
  corresponding to the non-normalizable zero-mode.
 To next order, we write
 \eq{\Psi_k(x)\sim \Psi_h(x) -k^2 e^{-U(x)/2}f(x).} Plugging this ansatz into Eq. \eqref{SMA3}, we get  \EQ{-U'(x)f'(x)+f''(x)=1.}{NextToLeadingOrder} The boundary conditions, Eq. \eqref{eqBC}, which apply for any \eq{k}, translate to  \eq{f(0)=f'(0)=0} and so a simple calculation yields  \EQ{f(x)=\int_0^xe^{U(x')}\int_0^{x'}e^{-U(x'')}\Intd x''\Intd x'.}{SMA4}
 The behavior of $f(x)$ for large $x$ can be analyzed as follows. Define $f_1(x)\equiv\int_0^{x}e^{-U(x')}\Intd x$. Then, for large $x$,
 \begin{align}
     f_1(x)&=\int_0^{x}(e^{-U(x')}-1+1)\Intd x \nonumber\\
     &= x + \int_0^{\infty}(e^{-U(x')}-1)\Intd x - \int_x^{\infty}(e^{-U(x')}-1)\Intd x \nonumber\\
     &\approx x + \ell_0 
 \end{align} 
 where we have defined the length $\ell_0 \equiv \int_0^{\infty}(e^{-U(x')}-1)\Intd x$ (note that \eq{\ell_0} is related to the second virial coefficient, see Sec. \ref{VirialTheorem125}). 
 Now, for large $x$,
 \begin{align}
     f(x)&=\int_0^{x}\left[e^{U(x')}f_1(x)-x-\ell_0 + x + \ell_0\right]\Intd x \nonumber\\
     &= \frac{x^2}{2} + \ell_0 x + \int_0^{\infty}\left[e^{U(x')}f_1(x)-x-\ell_0\right]\Intd x   \nonumber\\
       & \qquad\qquad {} - \int_x^\infty \left[e^{U(x')}f_1(x)-x-\ell_0\right]\Intd x \nonumber\\
     &\approx \frac{x^2}{2} + \ell_0 x + \mathcal{A}
 \label{AsyptoticsOffx} 
 \end{align}
 where the constant $\mathcal{A}$ with units of length${}^2$ is defined as $\mathcal{A}\equiv \int_0^{\infty}\left[e^{U(x')}f_1(x)-x-\ell_0\right]\Intd x.$ This behavior of $f$ can be seen to be consistent with the differential equation, Eq. \eqref{NextToLeadingOrder}.
In the matching region \textbf{II}, where \eq{1\ll x\ll1/k}, therefore:
\EQ{\Psi_{match}(x)\approx 1-k^2 ( x^2/2+ \ell_0 x + \mathcal{A})}{SMA7}

In region \textbf{III}, since \eq{x\gg1}, the \eq{U''(x)} and \eq{U'^2(x)} terms are negligible, and therefore, Eq. \eqref{SMA3} now reads 
\EQ{\frac{\partial^2}{\partial x^2}\Psi_k(x)\sim-k^2\Psi_k(x),}{SMA8} 
whose solution is \eq{\Psi_k(x)\approx A_k\cos(kx)+B_k\sin(kx)}. Comparing this solution with Eq. \eqref{SMA7} in the matching region, we find that \eq{A_k=1 - k^2\mathcal{A}} and \eq{B_k=-k\ell_0}.
We see that, if \eq{U(x)=0}, we have \eq{l_0 = 0}, and thus \eq{B_k=0}, namely in the force-free case the \eq{\sin(kx)} term is absent, as it should. We are now in a position to calculate the normalization \EQ{N_k^2 = \frac{2}{L(A_k^2+B_k^2)} \approx \frac{2}{L}(1+2\mathcal{A} k^2 - \ell_0^2 k^2).}{SMA8a}  It is interesting to note that the presence of the $B_k$  term, induced by the presence of $U(x)$,
results in a $O(1)$ leftward shift of $\ell_0$ in the original pure $\cos$ wave of the free particle case, in additional to the small change in normalization. This has a simple physical interpretation, which we will return to below.

\begin{figure}
    \centering
    \includegraphics[width=1.0\linewidth]{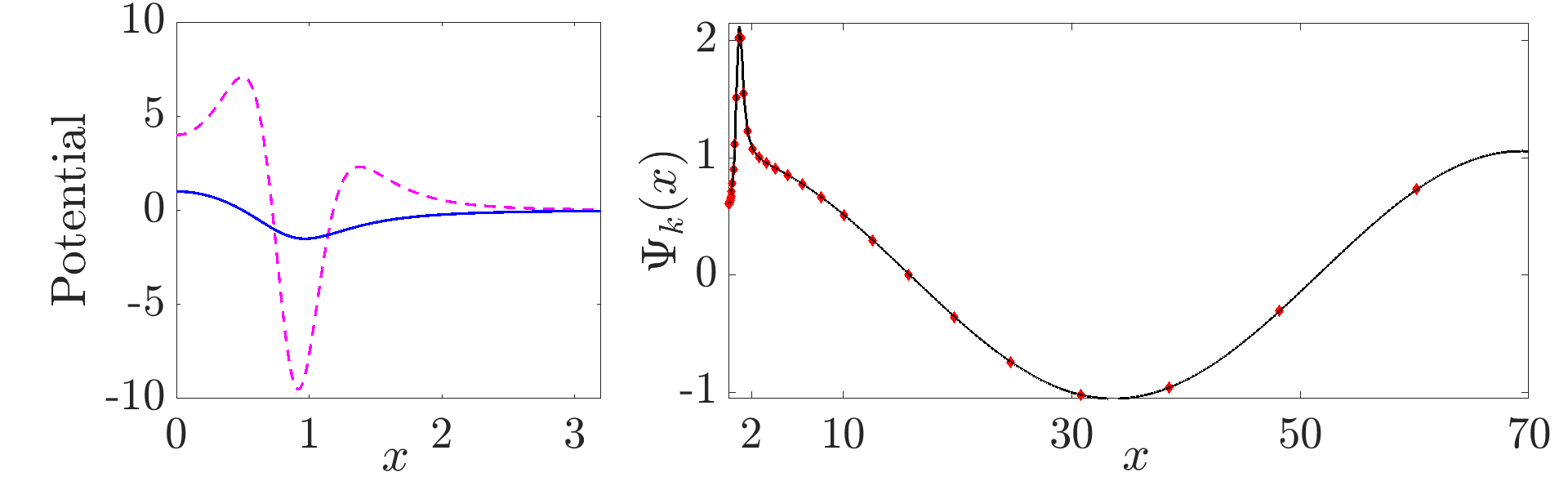}
    \caption{\footnotesize{(color online) Left panel: The potential \eq{U(x)=(1-4x^2)/(1+x^6)}, (blue line), {and} the effective potential in the Schr\"odinger Eq. \eqref{SMA3}: \eq{=-U''(x)/2+U'^2(x)/4} (magenta dashed line). Right panel: The eigenfunction \eq{\Psi_k(x)}, Eq. \eqref{WaveFunction} (red diamonds), {compared with} the exact numerical solution of the Schr\"odinger Eq. \eqref{SMA3} (black line), with the potential \eq{U(x)=(1-4x^2)/(1+x^6)} and \eq{k=0.088}. At small \eq{x}, the eigenfunction reflects the shape of the potential, since it is proportional to \eq{\exp[-U(x)/2]} (region \textbf{I}). At large \eq{x} (region \textbf{III}), the solution is proportional to \eq{A_k\cos(kx)+B_k\sin(kx)} (see Eq. \eqref{SMA8}).}}
    \label{PotentialsMathematica}
\end{figure}

A uniform approximation, for any \eq{x}, is seen to be: 
\begin{align}
\Psi_k(x)&\approx e^{-U(x)/2}\Big[\cos(kx)(1-k^2(\tilde{f}(x)+\mathcal{A}))\nonumber\\
  &\qquad\qquad\qquad {} -k\ell_0\sin(kx)\Big]
\label{WaveFunction}
\end{align} 
where we have defined  \EQ{\tilde{f}(x)\equiv f(x) - x^2/2 - \ell_0 x - \mathcal{A}.}{fTilde} 
In the left panel of Fig.~\ref{PotentialsMathematica}, we show the potential, \eq{U(x)=(1-4x^2)/(1+x^6)}, {and}  the effective potential in the ``quantum" problem, namely: \eq{-U''(x)/2+U'^2(x)/4}. In the right panel, we show {that} \eq{\Psi_k(x)} from Eq. \eqref{WaveFunction} {matches}  the exact numerical solution of the Schr\"odinger Eq. \eqref{SMA3}, with the above mentioned potential and \eq{k=0.088}. 

We are now in a position to take the $L\to\infty$ limit, wherein the sum over $n$ transforms into an integral over $k$,  \eq{\sum_k\rightarrow\int\Intd k \frac{L}{\pi}}.

For finite $x$, and $x_0$, then, using Eqs. \eqref{SMA2} and (\ref{SMA8a},\ref{WaveFunction}) the long-time density reads
\begin{align}
    P_t(x) &\approx e^{-U(x)} \int_0^\infty \frac{2\Intd k}{\pi} [1 - k^2(f(x)+f(x_0)-2\mathcal{A} + \ell_0^2)]\nonumber\\
    &\qquad\qquad\qquad \times e^{-D k^2 t}\nonumber\\
    &=   \frac{e^{-U(x)}}{\sqrt{\pi D t}} \left[1 - \frac{1}{2D t}(f(x)+f(x_0)-2\mathcal{A} + \ell_0^2)\right]
    \label{InnerCorrection}
\end{align}
 giving us the zeroth-order time-dependent Boltzmann-Gibbs factor, with a $1/t$ correction that grows quadratically in $x$.

\begin{figure}
    \centering
    \includegraphics[width=1.0\linewidth]{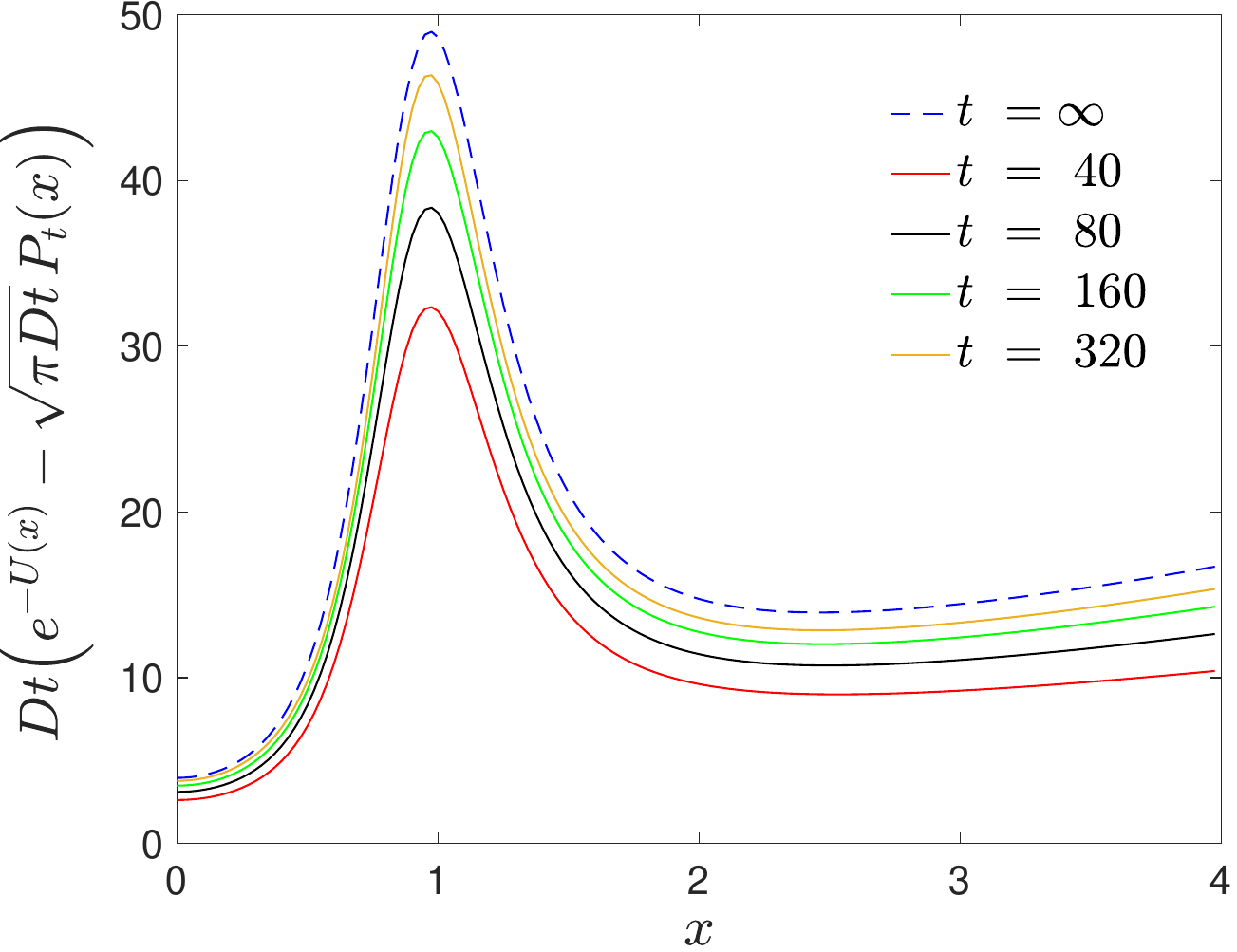}
    \caption{\footnotesize{The predicted scaled correction 
$Dt\left(e^{-U(x)} - \sqrt{\pi Dt}P_t(x)\right)$ {compared with}  the prediction from Eq. \eqref{InnerCorrection}, 
$\frac{1}{2}e^{-U(x)} (f(x)+f(x_0)-2\mathcal{A} + \ell_0^2)$, (labelled $t=\infty)$ for the case $U(x)= (1 - 4x^2)/(1+x^6)$ with $x_0 = 3$, for $t=40$, $80$, $160$, $320$. $D=1$. For this potential, $\ell_0=2.1719$, $\mathcal{A}=-5.3464$.}}
    \label{InnerCheck03}
\end{figure}

We test this prediction in Fig. \ref{InnerCheck03}, where we plot the scaled correction
$Dt\left(e^{U(x)} - \sqrt{\pi D t}P_t(x)\right)$ {and}  the prediction from Eq. \eqref{InnerCorrection}, 
$\frac{1}{2}e^{-U(x)} (f(x)+f(x_0)-2\mathcal{A} + \ell_0^2)$, for the case $U(x)= (1 - 4x^2)/(1+x^6)$ with $x_0 = 3$. We see that the numerics is converging to the prediction with increasing $t$, with the size of the correction growing as $x^2$. 

\begin{figure}
    \centering
    \includegraphics[width=1.0\linewidth]{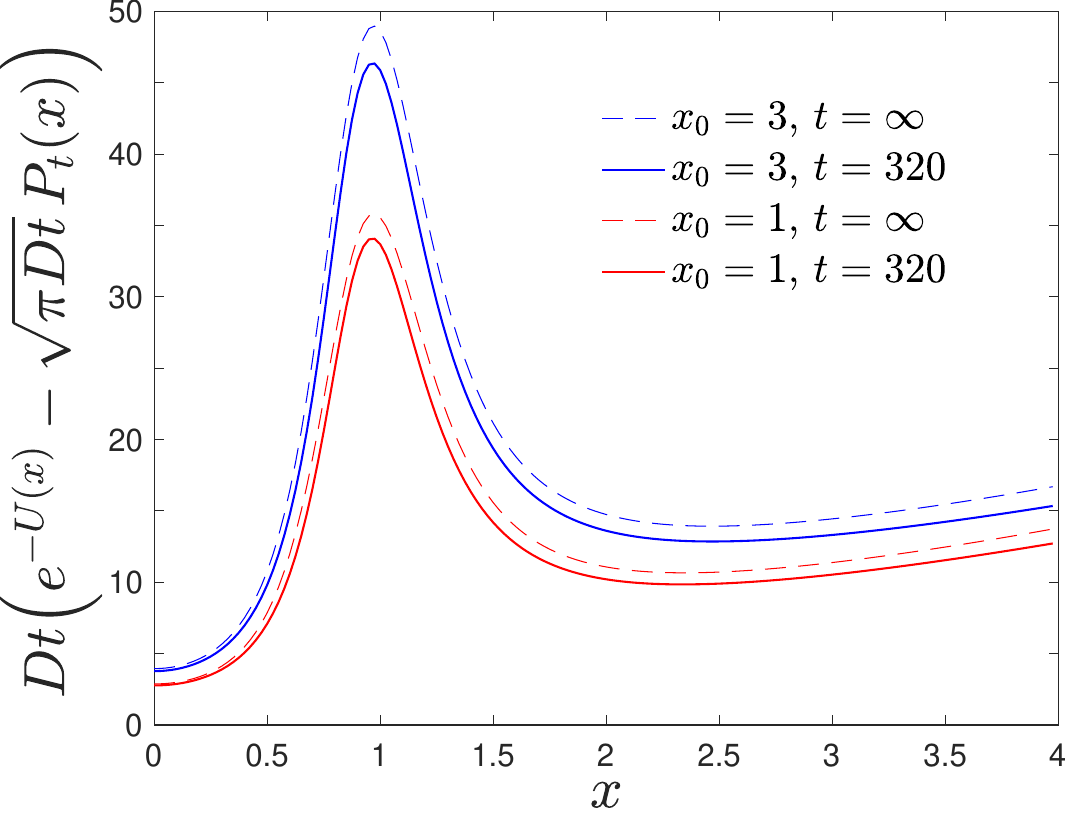}
    \caption{\footnotesize{The predicted scaled correction 
$Dt\left(e^{-U(x)} - \sqrt{\pi Dt}P_t(x)\right)$ {compared with}  the prediction from Eq. \eqref{InnerCorrection}, 
\eq{\frac{1}{2}e^{-U(x)} (f(x)+f(x_0)-2\mathcal{A} + \ell_0)}, (labelled $t=\infty)$ for the case $U(x)= (1 - 4x^2)/(1+x^6)$ with $x_0=1$ and $x_0 = 3$, for $t=320$. $D=1$, $\ell_0=2.1719$, $\mathcal{A}=-5.3464$.}}
    \label{InnerCheck13}
\end{figure}

To test the dependence on $x_0$, we plot in Fig. \ref{InnerCheck13}, the predicted correction, {and}  the simulation {results} at $t=320$ for the same potential, for both $x_0=1$ and $x_0=3$. The formula is seen to correctly capture the $x_0$ dependence, with the correction larger in magnitude for larger $x_0$,

\begin{figure}
    \centering
    \includegraphics[width=1.0\linewidth]{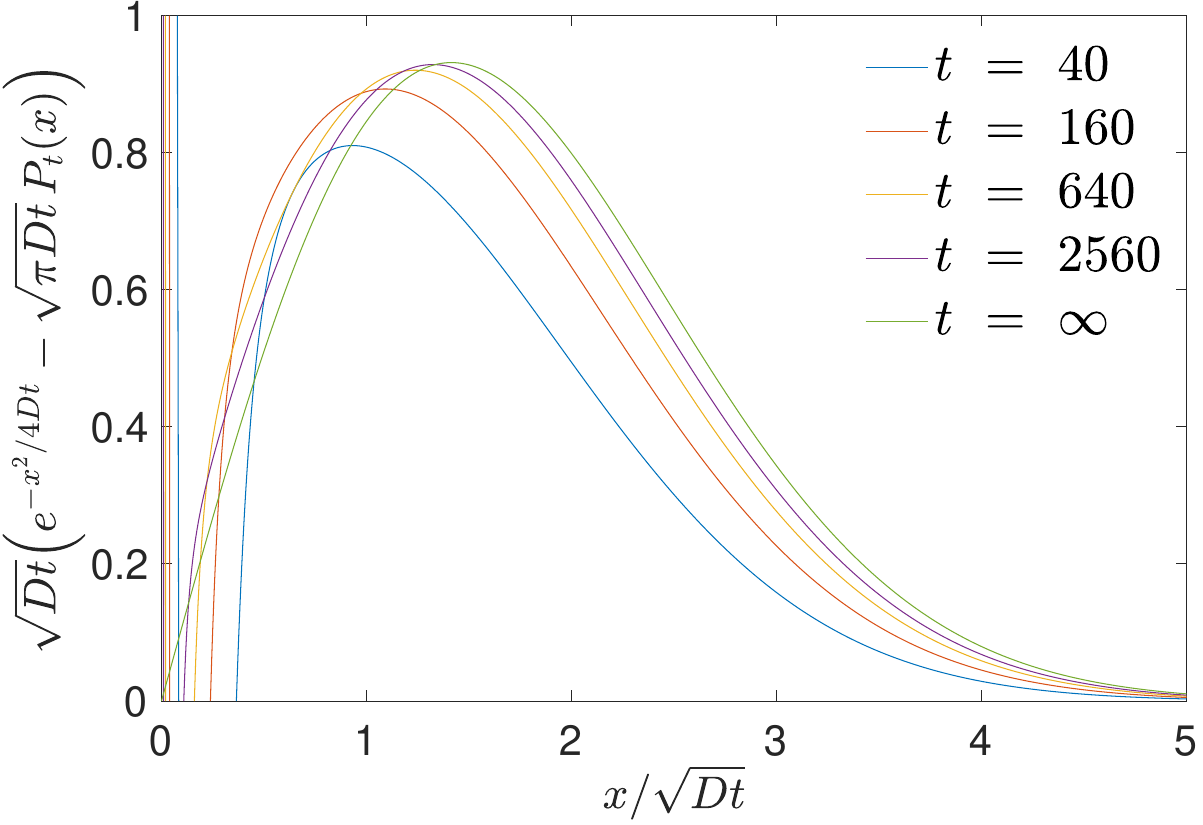}
    \caption{\footnotesize{The predicted scaled correction in the outer regime, $x\sim \sqrt{Dt}$,
$\sqrt{Dt}(e^{-x^2/4Dt} - \sqrt{\pi D t}P_t(x))$ {compared with}  the prediction from Eq. \eqref{OuterCorrection1}, 
\eq{\frac{\ell_0 x}{2}(-\mathcal{A} + \ell_0^2 + f(x_0))e^{-x^2/4Dt}} , (labelled $t=\infty)$ for the case $U(x)= (1 - 4x^2)/(1+x^6)$ with  $x_0 = 1$, for $t=40$, $160$, $640$, and $2560$. $D=1$, $\ell_0=2.1719$. }}
    \label{OuterCheck1first}
\end{figure}

\begin{figure}
    \centering
    \includegraphics[width=1.0\linewidth]{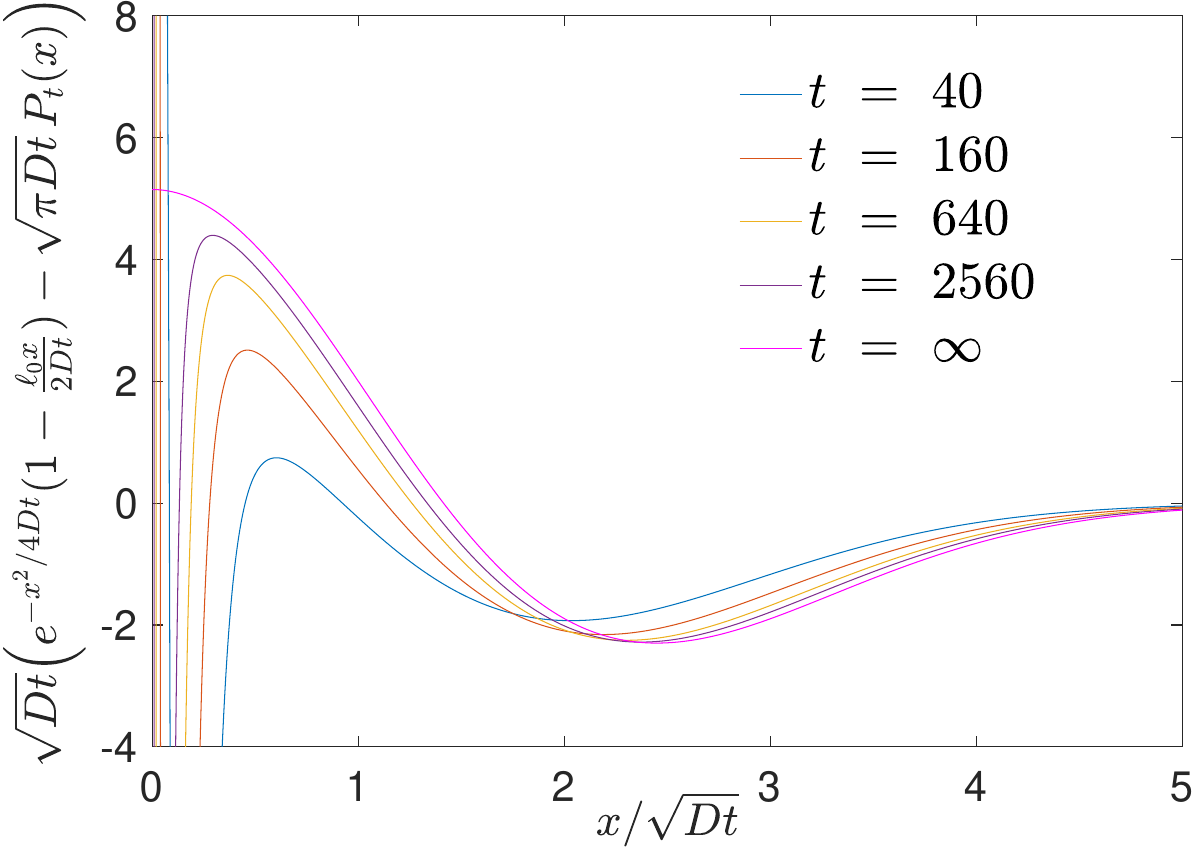}
    \caption{\footnotesize{The predicted scaled second order correction in the outer regime, $x\sim \sqrt{Dt}$,
$\sqrt{Dt}(e^{-x^2/4Dt}\left(1-\frac{\ell_0 x}{2Dt}\right) - \sqrt{\pi D t}P_t(x))$ {compared with}  the prediction from Eq. \eqref{OuterCorrection}, 
\eq{\frac{2Dt-x^2}{4D^2 t^2}e^{-x^2/4Dt}} , (labelled $t=\infty)$ for the case $U(x)= (1 - 4x^2)/(1+x^6)$ with  $x_0 = 1$, for $t=40$, $160$, $640$, and $2560$. $D=1$, $\ell_0=2.1719$, $\mathcal{A}=-5.3464$. }}
    \label{OuterCheck12nd}
\end{figure}

For large $x$, of order $\sqrt{t}$, but $x_0$ still $O(1)$, the long-time density is
\begin{align}
    P_t(x) &\approx  \int_0^\infty \frac{2\Intd k}{\pi} [(1 - k^2 \mathcal{A})\cos(kx) - k\ell_0 \sin(kx)]\nonumber\\
    &\qquad \times [1 - k^2 f(x_0) ][1 + 2k^2\mathcal{A} - k^2\ell_0^2 ] e^{-Dk^2 t}\nonumber\\
    &=   \frac{e^{-x^2/4Dt}}{\sqrt{\pi D t}} \Bigg[1 - \frac{2Dt-x^2}{4D^2 t^2}(-\mathcal{A} + \ell_0^2 + f(x_0))\nonumber\\
    &\qquad\qquad \qquad\qquad{} - \frac{\ell_0 x}{2Dt}\Bigg]
    \label{OuterCorrection}
\end{align}
Thus, it turns out that in this regime, the leading order correction to $P_t$ comes from a rightward  shift in the Gaussian by an amount $\ell_0$, due to the shift in the phase of $\Psi_k$ discussed above.  Thus, at large distances, the position of the diffusive source is effectively at $x=-\ell_0$. 
As this shift leads to an O($1/\sqrt{t}$) relative change in the solution, if we consider just this leading change, the O($1/t$) terms are negligible, and the solution simplifies to
\EQ{
    P_t(x) 
    \approx  \frac{e^{-x^2/4Dt}}{\sqrt{\pi D t}} \Big[1 
     - \frac{\ell_0 x}{2Dt}\Big].}{OuterCorrection1}
     Note that $\ell_0$  may take either positive or negative values hence the sign of the correction term depends on the force field (see Sec. \ref{VirialTheorem125}, where we relate $\ell_0$ to the virial theorem). 
     This prediction is tested in Fig. \ref{OuterCheck1first}, where we see very good agreement. In Fig. \ref{OuterCheck12nd}, we check the validity of the $1/t$ relative change via  the difference of the simulation to the first order outer solution, Eq. \eqref{OuterCorrection1}, and see that here too the agreement is excellent.
     
More generally, for arbitrary $x$, we have the uniform solution
\begin{align}
P_x(t)&\approx \frac{e^{-(x+\ell_0)^2/4Dt}e^{-U(x)}}{\sqrt{\pi D t}}   \nonumber\\
&\qquad\times \Bigg[1-\frac{2Dt-x^2}{4D^2 t^2}(-\mathcal{A} + \ell_0^2/2 + f(x_0) + \tilde{f}(x)) \Bigg], 
\label{UNIFORM}
\end{align}
where \eq{\tilde{f}(x)} is defined in Eq. \eqref{fTilde}. In Sec. \ref{NextToLeadingOrderBehaviorOfTheMSD}, we use the next-to leading order behavior of \eq{P_t(x)} in order to obtain a correction term to the leading-order, linear behavior of the mean-squared displacement \eq{\langle x^2\rangle}.  


{\em Comment on normalization.}
We can verify that the solution in Eq. \eqref{UNIFORM} is normalized to unity, in the following way: Integrating \eq{\exp[-U(x)-x^2/(4Dt)]}, we find that \eq{\int_0^\infty e^{-U(x)-\frac{x^2}{4Dt}}\Intd x\approx \sqrt{\pi Dt}+\int_0^\infty (e^{-U(x)}-1)e^{-\frac{x^2}{4Dt}}\Intd x,} from which, for potentials that fall-off faster than \eq{1/x^2} at large \eq{x}, in the long-time limit we get \eq{\int_0^\infty e^{-U(x)-x^2/4Dt}\Intd x\approx\sqrt{\pi Dt}+\ell_0}.  So, the first term  of $\int_0^\infty P_t(x)\Intd x$ is approximately $1+\ell_0/\sqrt{\pi D t}$. Similarly, the $\ell_0x/2Dt$ term in Eq. \eqref{OuterCorrection1},  yields \eq{-\int_0^\infty{\ell_0xe^{-x^2/(4Dt)}\Intd x/[2\sqrt{\pi}(Dt)^{3/2}]}}.  This cancels out the correction to the normalization from the leading term. 

\section{Time and ensemble-averages}
\label{SecTimeAndEnsembleMeans}
Let us now focus on the limit of long times, where the correction terms to the leading-order behavior of \eq{P_t(x)} are negligible with respect to the uniform approximation, Eq. \eqref{UniformApproximation}.
 To define the long-time limit of averages, we distinguish between two types of observables:
integrable and non-integrable observables, with respect to the non-normalized Boltzmann state.
We consider first the indicator function
\begin{equation} 
{\cal O}[x(t)] = I(x_1 <x(t)<x_2),
\label{eqTA01}
\end{equation}
where $I(\cdots)=1$ if the condition in the parenthesis is satisfied, and zero otherwise. Along the trajectory \eq{x(t)} of the particle, this observable, \eq{I(\cdot)}, switches between values $+1$ and $0$,
corresponding to whether the particle is present in the domain $(x_1,x_2)$ or not. 
Here, $x_1$ and $x_2$ are the experimentalist's matter of choice. 

 The ensemble-average of this observable, which in principle can be obtained
from a packet  of non-interacting particles, at some time $t$ is
\begin{align}
\langle I (x_1<x(t) <x_2)\rangle &= \int_0 ^\infty I(x_1<x <x_2) P_t(x) {\rm d} x\nonumber\\&\sim 
\int_{x_1}  ^{x_2} e^{ - \beta V(x) } / \mathcal{Z}_t. 
\label{eqTA02}
\end{align}
This result is valid in the limit of long times, when $x_1$ and $x_2$ are much smaller
than the diffusion length-scale $l(t)=\sqrt{4 D t}$, namely we used the approximation
$\exp[-(x_2)^2/ 4 D t] \simeq 1$. We see that, while the Boltzmann
factor $\exp( - \beta V(x))$ is not normalized, it is used to obtain the ensemble
averages. In this case the observable is zero at large distances, hence this observable
cures the non-integrability of the infinite density. More generally, for observables integrable
with respect to the non-normalized Boltzmann factor we have, 
using 
Eq. (\ref{UniformApproximation})
\begin{equation}
\lim_{t \to \infty}  \mathcal{Z}_t \langle {\cal O}(x) \rangle = \int_0 ^\infty \exp[ -\beta  V(x)] {\cal O}(x) {\rm d} x.
\label{eqTA03}
\end{equation}
Eqs. (\ref{eqTA02},\ref{eqTA03}) are valid also for the case when the system reaches a steady state,
and then $\mathcal{Z}_t$ is the normalizing partition function; in that case Eq. (\ref{eqTA02}) is simply the
probability of finding the particle in thermal equilibrium in the interval $(x_1,x_2)$. 

 The time that a particle spends in the domain $(x_1,x_2)$ is called the residence time or the occupation
time, and it is denoted $t_{x_1<x<x_2}$. This variable fluctuates from one trajectory to another,
however when the system reaches a steady state (i.e. if the potential is  binding),
the occupation fraction in the long measurement-time limit
clearly satisfies  $\lim_{t \to \infty} t_{x_1<x<x_2}/t= \mbox{Prob}(x_1<x<x_2)$ and the latter is obtained from the Boltzmann
measure  
\begin{equation}
 \lim_{t \to \infty} t_{x_1<x<x_2}/t= 
{\int_{x_1} ^{x_2} e^{- \beta  V(x)} {\rm d} x \over Z}.
\label{eqTA04}
\end{equation}
This result can be obtained also from Birkhoff's ergodic hypothesis in standard Boltzmann-Gibbs theory.

What is the corresponding behavior for weakly binding potentials where infinite ergodic theory is relevant? 
The observable $t_{x_1<x<x_2}/t$ is the finite-time average
\begin{equation}
{ t_{x_1 <x<x_2} \over t} = {\int_0 ^t I (x_1<x(t)<x_2) {\rm d} t \over t}.
\label{eqTA05}
\end{equation}
Let us first consider the ensemble-average of this observable, which is obtained by averaging over an
ensemble of paths, each trajectory yielding its own residence time. Here we have 
\begin{equation}
\left\langle { t_{x_1 <x<x_2} \over t} \right\rangle 
= {\langle  \int_0 ^t I(x_1<x(t) <x_2) {\rm d} t \rangle  \over t}.
\label{eqTA06}
\end{equation}
To calculate this value we can switch the order of the ensemble-averaging procedure with the time integration, and use
 $\langle I (x_1<x(t)<x_2)\rangle = \int_{x_1} ^{x_2}
 P_t(x)  {\rm d} x$.  Now, we need to perform the time integration, however considering the long-time
limit (and neglecting short-time effects), this calculation is straight forward: 
using $\mathcal{Z}_t = \sqrt{ \pi D t}$, Eq. 
(\ref{eq19}), we get
\begin{align}
\left\langle { t_{x_1 <x<x_2} \over t} \right\rangle &\sim 
{1 \over t} \int_0 ^t {\rm d} t {\int_{x_1} ^{x_2} e^{ - \beta V(x)} {\rm d} x \over \mathcal{Z}_t } = \nonumber\\
& 2{ \int_{x_1} ^{x_2} \exp[- \beta V(x)]{\rm d} x \over \mathcal{Z}_t}.
\label{eqTA062}
\end{align}
The factor $2$ is a consequence of the time integration, since $\int_0 ^t t^{-1/2} {\rm d} t= 2 t^{1/2}$,
and note that  we may take here the lower limit of the integration to zero, without any affect on the long-time limit.

As for the indicator function, now consider the averaged potential energy, with the uniform approximation, Eq. \eqref{UniformApproximation}:
\begin{equation}
\langle V(x) \rangle \sim { \int_0 ^{\infty} \exp( -\beta  V(x) - x^2/ 4 D t) V(x) {\rm d} x 
\over \mathcal{Z}_t}.  
\label{eqTA07}
\end{equation}
In the long-time limit, we have
\begin{equation}
\lim_{t \to \infty} \mathcal{Z}_t \langle V(x) \rangle = \int_0 ^\infty \exp( - \beta V(x) ) V(x) {\rm d} x, 
\label{eqTA08}
\end{equation}
since the potential is zero beyond some length-scale $l_1$, and $\exp( - x^2 /4 D t) \sim 1$ 
for $0<x<l_1$. The total potential energy is decreasing with time (in absolute value, and in contrast with the
 entropy which is increasing), as particles are
escaping the well, traveling to the bulk and exploring the spatial domain where the force is negligible. Here, it is important to note
that the process is recurrent, so any particle which escapes the surface to any distance as long as we wish,
will eventually return to the regime of non-zero potential with probability one (to the local minimum of  the Lennard-Jones potential, for example). This means that if we perform an
experiment with $N\gg1$ particles, it is more likely to find them in the medium, beyond
$l_1$, after some finite time. Still, since one always observes the return of the particles,
there is always a non-negligible number of them which are residing in the vicinity of the surface (no escape is forever). 
%
%

Now, consider the ensemble-average of the deterministic part of the force field $f(x) = - \partial_x V(x)$

\begin{align}
    \langle f(x)\rangle&=k_BT\frac{\int_0 ^\infty \partial_x e^{ - V(x)/k_B T} \Intd x }{\mathcal{Z}_t}= \nonumber\\ 
& \frac{k_B T}{\mathcal{Z}_t} \left\{  \exp[ - \beta V(\infty)] - \exp[ -\beta  V(0)] \right\}.
\end{align}
Since this observable is also integrable with respect to the infinite density, Eq. \eqref{EqID5}, we get $\langle f(x) \rangle = k_b T /\mathcal{Z}_t$ or 
\begin{equation}
\lim_{t \to \infty} \sqrt{\pi Dt} \langle f(x) \rangle = k_B T, 
\label{eqTA11}
\end{equation}
if we consider the case of the one-sided system, and \eq{=0.5k_BT} in the two-sided case (since \eq{\mathcal{Z}_t\rightarrow2\mathcal{Z}_t}). 
Notice that this limit does not depend on the specific shape of the potential. 

For all the integrable observables above (and in fact for any integrable observable), we 
find  a connection between the time and ensemble-average, which is a generalization of
the Birkhoff law from standard thermodynamics{, }
namely the doubling effect seen in Eq. \eqref{eqTA062} is a general feature for this class of observables. Consider an observable
${\cal O}[x(t)]$ which is integrable with respect to the non-normalized Boltzmann factor,  
then the ensemble-average of the time-average is 
\begin{equation}
\langle \overline{{\cal O}[x(t)]} \rangle = 2 \langle {\cal O}(x) \rangle
\label{eqTA12}
\end{equation}
where
\begin{equation}
\langle {\cal O}(x) \rangle = {\int_0 ^\infty {\cal O}(x)  \exp[ -\beta  V(x)] {\rm d} x \over \mathcal{Z}_t}.
\label{eqTA13}
\end{equation}
The factor $2$ is a consequence of the diffusive nature of the process, which leads to the integration over the time-dependent partition function, hence this doubling
effect might be widely observed. {The numerical results  which support Eq. \eqref{eqTA12}, were presented in our previous work, Ref. \cite{aghion2019non}, where we showed that the simulations agreed with the theory. Below, in Sec. \ref{ErgodicityIndDimensions}, we show numerical evidence for a variation of Eq. \eqref{eqTA12}, which is valid also in dimensions \eq{d\geq1}.}

\section{Fluctuations of the time-averages}
\label{FluctuationsOfTimeAverages}

\begin{figure}[t]
\includegraphics[width=1.0\linewidth]{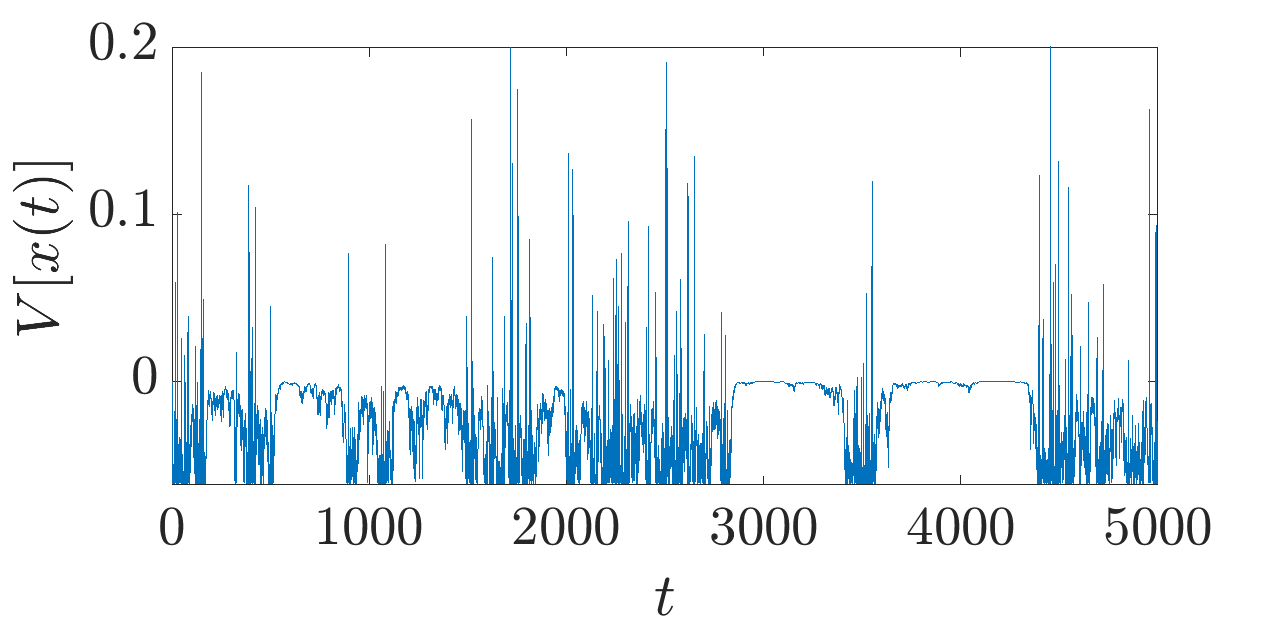} 
\includegraphics[width=1.0\linewidth]{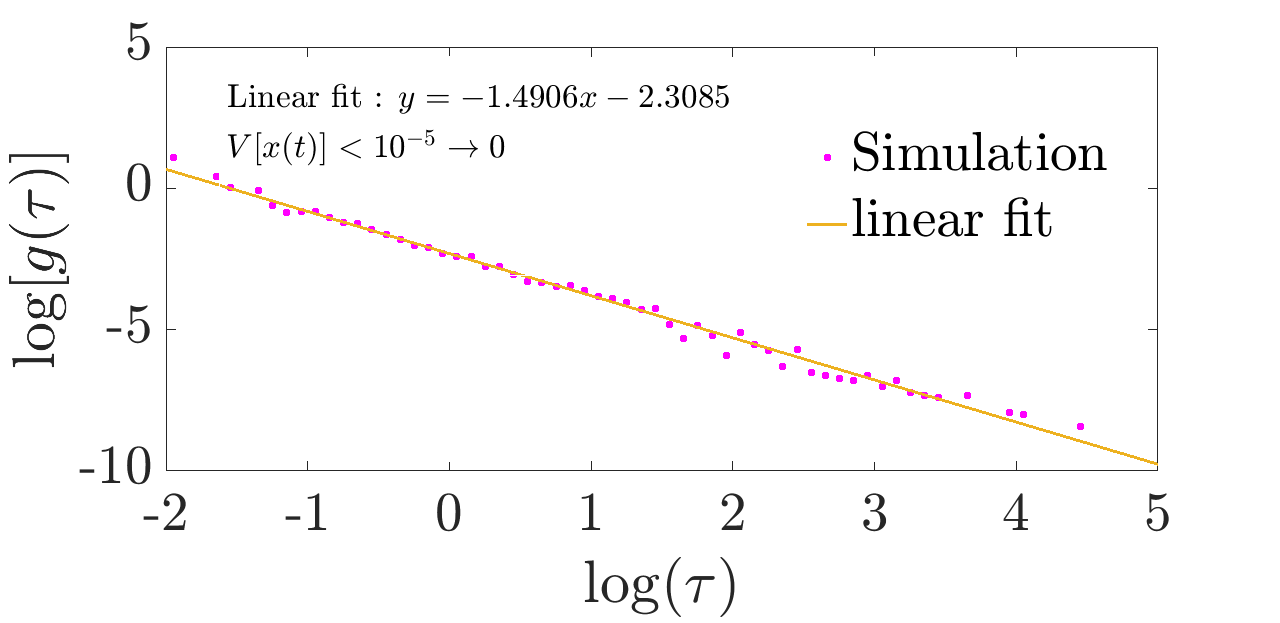} 
    \caption{{\footnotesize{\textit{Upper panel}: Simulated time series of the potential energy of a particle in the Lennard-Jones potential Eq. \eqref{eq09}. It exhibits long periods during which \eq{|V[x(t)]|<\epsilon}, where  \eq{\epsilon>0} can be as small as we wish, and short periods where the absolute value of the energy is high. Small absolute values of the potential energy corresponds to events where the particle strays far away from the potential minimum (the particle being in the range \eq{x\gg1}). \textit{Lower panel}: The probability distribution \eq{g(\tau)} of the durations (\eq{\tau}s) of the time periods where the absolute value of the energy is lower than \eq{\epsilon=10^{-5}} (here, the total measurement time was \eq{t=2\times10^6}). \eq{\log[g(\tau)]} versus \eq{\log(\tau)}, obtained from the simulation (magenta circles), is matched with a linear fit (orange line) with a slope of \eq{\approx -3/2}, which implies that \eq{g(\tau)} has a power-law shape similar to free Brownian motion.}}}
\label{SM125}
\end{figure} 

 The time-average $\overline{{\cal }O[x(t)]}$ in the time-independent
 canonical setting is equal to the ensemble-average, in the long-time limit (ergodicity).  
In our case, the time-averages fluctuate between different trajectories,
which is a common theme in single-molecule experiments,
and here we explore the fluctuations. 
To start, we again consider  the indicator function, 
${\cal O}[x(t)]=I(x_1<x(t)<x_2)$, defined in 
Eq. (\ref{eqTA01}).
However, our results  are far more general than that. As we will show, 
the fluctuations of time-averages
of  observables integrable with respect to the 
non-normalized Boltzmann density follow a universal law, in the spirit
of the Aaronson-Darling-Kac theorem \cite{aaronson1997introduction}.  

For simplicity, let us consider $x_1=0$.
The process $I(x_1<x(t)<x_2)$, starting inside the region $(0,x_2)$, is switching randomly between
two states, with sojourn times in the interval close to the surface
denoted $\tau^{{\rm in}}$, when $I(x_1<x(t)<x_2)=1$, and $\tau^{{\rm out}}$, when $I(x_1<x(t)<x_2)=0$. The first time interval in the domain $(0,x_2)$
is denoted $\tau_1 ^{{\rm in}}$, and the rest follow,
so the sojourn times in the two states are given by
the sequence
\begin{equation}
\left\{ \tau^{{\rm in}} _1, \tau^{{\rm out}} _1,
 \tau^{{\rm in}} _2, \tau^{{\rm out}} _2, \cdots \right\}.
\label{eqFLT01}
\end{equation}
These times {can be treated as}  mutually independent, identically distributed random variables{, since temporal correlations in the Langevin Eq. \eqref{eq08} decrease exponentially fast in time}. We denote the probability density functions
 of \textit{out} and \textit{in}
 sojourn times with $\psi_{{\rm out/in}}(\tau)$, respectively.
Importantly, in the long-time limit; 
$\psi_{{\rm out}}(\tau)\propto \tau^{-3/2}$. This well-known result is related of course to the flatness of the
potential field at large \eq{x}. In this regime, the process
$x(t)$ is  controlled by
diffusion and while it is recurrent, so the density
$\psi_{{\rm out}}(\tau)$ is normalized, the average sojourn time
in the {\em out} state diverges. 
This absence of a typical timescale,  together with the diverging partition function, are precisely the reasons for the  failure of the standard (Birkhoff) ergodic theory, and the emergence of Boltzmann-like infinite ergodic theory.  
The sojourn times in the {\em in} state are thinly distributed and, importantly, the moments of $\tau^{\textrm{in}}$ are finite, which is clearly the case since the interval $(0,x_2)$ is of finite length. 

 Let $n$ be the number of switching events from the  {\em in} to 
the {\em out} states.
For a fixed measurement time $t$, this number is random. We claim that 
the distribution of $n$ is determined by the statistics
of the  ${  out}$  times,
when $t$ is large. Roughly speaking, the ${ in}$ sojourn times are very short when compared to  the ${ out}$ times, since those
have an infinite mean. This means that in the time interval $(0,t)$, we will typically 
observe an ${ out}$ sojourn time of the order of magnitude of the measurement time
$t$, and the size of this largest interval  controls the number of \eq{out}-to-\eq{in} transitions 
(if the largest {\em out} time is very long, then $n$ is small, compared with a realization with a shorter maximal {\em out} time). 

 Note that the time-average of $I(x_1<x(t)<x_2)$ is equal to the sum of 
the {\em in} times, 
divided by the measurement time;
$\overline{I} = \sum_{i=1} ^n \tau_i ^{{\rm in}}/t$, where we assume, without any loss of generality,
that at time $t$ the process is in state {\em out}.  
The mean $\langle \overline{I}\rangle$ was already obtained rigorously
from the non-normalized density in Eq. \eqref{eqTA062}, but with the notations of renewal theory
we have $\langle \overline{I} \rangle \simeq \langle \tau^{{\rm in}} \rangle \langle n \rangle /t$ and since $\langle n \rangle \sim t^{1/2}$ we have
$\langle \overline{I} \rangle \propto t^{-1/2}$, as we found earlier.  
The behavior $\langle n \rangle \propto t^{1/2}$  is well known in renewal theory,
and with a hand-waving argument we note that the 
effective average sojourn time 
is $\langle \tau_{{\rm eff}} \rangle =
\int_0 ^t \tau \psi_{{\rm out}} (\tau) {\rm d} \tau \propto t^{1/2}$,
and hence
$\langle n \rangle \sim t /\langle \tau_{{\rm eff}} \rangle \propto t^{1/2}$.

 Let us now consider the second moment of the time-average. {Since, for any \eq{i\neq j}, \eq{\langle (\tau^{\rm in}_i)^2\rangle=\langle (\tau^{\rm in}_j)^2\rangle} and \eq{\langle \tau^{\rm in}_i\tau^{\rm in}_j\rangle=\langle \tau^{\rm in}_i\rangle^2}, we argue that}   
\begin{align}
\langle \overline{I}^2  \rangle &= {\langle\left( \sum_{i=1} ^n \tau_i ^{{\rm in}} \right)^2 \rangle \over t^2 } \nonumber\\
&= { \langle n \rangle \langle (\tau^{{\rm in}})^2 \rangle
+ \langle n(n-1) \rangle \langle \tau^{{\rm in}} \rangle^2 \over t^2} .
\label{eqFTA02}
\end{align}
 Now we are ready to get to the main point of this section. 
Considering the variance of the time-average 
\begin{align}
 &\langle (\overline{I})^2 \rangle   - \langle \overline{I}\rangle^2=
{\langle (t_r)^2 \rangle  - \langle t_r\rangle^2 \over t^2} = 
\nonumber\\
&{ \langle n^2 \rangle - \langle n \rangle^2 \over t^2}  \langle \tau^{{\rm in}}\rangle^2 +
{\langle n \rangle \over t^2}  \underbrace{\left[ \langle (\tau^{{\rm in}})^2\rangle
- \langle \tau^{{\rm in}}\rangle^2 \right]}_{\textrm{Var}(\tau^{\rm in})}, 
\label{eqFTA03}
\end{align}
we see that the second term is negligible, compared with the first,
since from renewal theory we know that $n \sim t^{1/2}$.
This means that the fluctuations of $\tau_{{\rm in}}$ are irrelevant,
and there is a single important timescale describing the process,
which is the average sojourn time $\langle \tau^{{\rm in}} \rangle$.
We now normalize the variance using
$\langle \overline{I} \rangle \sim \langle n \rangle \langle \tau^{{\rm in}}\rangle /t$ and find
\begin{equation}
{\langle \overline{I}^2 \rangle - \langle \overline{I}\rangle^2\over \langle \overline{I} \rangle^2 }\rightarrow {\langle n^2 \rangle - \langle n \rangle^2 \over \langle n \rangle^2}.
\label{eqFTA04}
\end{equation}
This analysis can be continued to higher  order moments and it yields 
\begin{equation}
{\overline{I} \over \langle \overline{I} \rangle } \rightarrow { n \over \langle n \rangle} \equiv \eta.
\label{eqFTA05}
\end{equation}
This means that
the residence time in $(x_1,x_2)$, divided by the mean residence time,
is equal in a statistical sense to number of the {\em{in}}-to-{\em{out}} transitions over their mean. 

 The probability density function of $0<\eta$ is known from renewal theory \cite{godreche2001statistics},
and since $\psi_{{\rm out}} (\tau) \propto \tau^{-3/2}$ we find
\begin{equation}
\mbox{PDF} (\eta) = {2 \over \pi} e^{ -\eta^2/\pi}.
\label{eqFTA06}
\end{equation}
This has, by definition, unit mean. Naively, the reader might be tempted to believe that this result is related to the Gaussian central limit theorem. However,
this is not the case, since for  sojourn-time distributions
 with other fat tails 
 we will get a form very different from Gaussian \cite{he2008random,akimoto2015distributional,aghion2019non}. In-fact, \eq{\mbox{PDF} (\eta)} {is a special case of a more general density function} known as the Mittag-Leffler distribution, which in turn is related
to L\'evy statistics. 
{Note, that the Aaronson-Darlin-Kac theorem \cite{aaronson1997introduction} predicts that the distribution of the time average of a process with an infinite measure, will be given by the Mittag-Leffler distribution in the form of $
 \mathscr{M}_\alpha(\eta)= \frac{\Gamma^{1/\alpha}(1+\alpha)}{\alpha \eta^{1+1/\alpha}}l_{\alpha,0}\left[\frac{\Gamma^{1/\alpha}(1+\alpha)}{ \eta^{1/\alpha}}\right]$, with \eq{l_{\alpha,0}(\#)} being the one-sided L\'evy density (defined as the inverse-Laplace 
transform of \eq{\exp(-u^\alpha)}, from \eq{u \rightarrow \#}), see e.g., \cite{korabel2009pesin,akimoto2013aging,radice2020statistics}. The exponent \eq{\alpha}, we argue, is determined by the first return probability \eq{\psi_{{\rm out}} (\tau) \propto \tau^{-1-\alpha}}, and in our case, \eq{\alpha=1/2} means that \eq{\mathscr{M}_{1/2}(\eta)} is equal to Eq. \eqref{eqFTA06}. Other values of \eq{\alpha} are found in the case of diffusion in logarithmic potentials, as explained in Ref. \cite{aghion2019non} (see also \cite{dechant2011solution}), which are out of the scope of this manuscript, but in that case a derivation of the Mittag-Leffler distribution can also be made following the same lines as presented below.}

 A hand-waving argument for Eq. (\ref{eqFTA06}), works as follows: 
Consider $n$ independent, identically distributed random variables ${\tau^{\textrm{out}}_1, ... \tau^{\textrm{out}}_n}$, which correspond
in our physical model to  the times in the state \eq{out}. According to the L\'evy central limit theorem \cite{metzler2000random}, the probability density function (PDF) of these
times is the one-sided L\'evy density with index $1/2$ 
\begin{equation}
l_{1/2,0} (\tau) = { 1 \over 2 \sqrt{\pi}} \tau^{-3/2} \exp\left( - {1\over 4 \tau}\right).
\label{eqllll}
\end{equation}
Here, like $\psi_{{\rm out}}(\tau)$, $l_{1/2,0}(\tau) \propto \tau^{ - 3/2}$
and this fat-tailed behavior allows us to consider a specific choice of the \eq{out}
times distribution, in the sense that asymptotically the results are not
sensitive to the short $\tau$ behavior of $\psi_{\textrm{out}}(\tau)$. 
We use  dimensionless units, 
and since eventually we consider
the dimensionless variable $\eta$, this is not a problem. 
The Laplace transform \eq{LT[f(\tau)]=\int_0^\infty f(\tau)\exp(-\tau u)\Intd \tau} of Eq. (\ref{eqllll}) is $\exp( - u^{1/2})$.
Now, consider the random variable $y = \sum_{i=1} ^n \tau_i / n^2$.
The PDF of $y$ is also the one-sided stable law Eq. (\ref{eqllll}),
 since
it is easy to check  that $\langle \exp( - u y ) \rangle= \exp( - u^{1/2})$.
We are interested in the probability distribution of $n$, and we fix the measurement
time $t$ to be the sum of the sojourn times $t = \sum_{i=1} ^{n} \tau_i$.
Hence $y = t/n^2$, and 
$$ \mbox{PDF}(n) = \mbox{PDF}(y) |{{\rm d} y \over {\rm d} n}|= $$
\begin{equation}
l_{1/2,0}\left({t \over n^2}\right)| { 2 t \over n^3}|=
{1 \over \sqrt{ \pi t} } \exp\left( - { n^2 \over 4 t}\right).
\label{eqllll1}
\end{equation}
Thus the density of $n$ is half a Gaussian. From here, we find 
$\langle n \rangle = 2 \sqrt{t/\pi}$, and switching to the random variable
$\eta= n/\langle n \rangle$ we get Eq.  
(\ref{eqFTA06}). Throughout this derivation, we treat $n$ as a continuous variable, which
makes sense in the long-time limit, and 
can be justified using well-known rigorous results.

 We note that,
mathematically, the number of switching events $n$ is formally infinite.
This is related to the fact that the Langevin
 trajectories are continuous, hence once
we have one transition, we experience many of them. This is not
a major problem since  we actually considered the scaled
random variable $\eta=n/\langle n \rangle$ which has a unit mean.
To put it differently,
since $\langle n \rangle \sim t^{1/2}$ in the long-time limit,
we consider a scaled variable which is perfectly well behaved. 
From the measurement point of view,
we sample the trajectory with a finite rate, so $n$ is always finite,
and this is also true in simulations, where we use discrete steps in space and time
(in the limit of large \eq{n}, the results will not be sensitive to the sampling 
rate
and the discretization).

 Inspired by infinite ergodic theory, we claim that the  Mittag-Leffler distribution of time-averages is a far more general result. For example, consider the time-average of the potential energy 
$\overline{V}$. Also here the observable $V[x(t)]$ is switching between 
long periods where it is nearly zero (when the particle is far in the bulk), to relatively short bursts when this observable is nonzero,
when the particle is close to the surface. Again the statistics
of the number of times that the particle visits the domain where 
$V(x)$ is non-negligible, is similar to that of $n$, and its statistics is controlled by the
first-passage probability density function from the bulk to the vicinity 
of the wall. Again, the latter is the fat-tailed density with the familiar
 $\tau^{-3/2}$ law that we have just
mentioned above.  So we have
\begin{equation}
{\overline{{\cal O}[x(t)]} \over \langle \overline{{\cal O}[x(t)]} \rangle } \rightarrow
 \eta,
\label{eqFTA07}
\end{equation}
and Eq. (\ref{eqFTA06}) still holds. Note that this yields a complete
description of the problem in the sense that $\langle \overline{{\cal O}}\rangle$ is calculated in principle 
 with the non-normalized Boltzmann  density, and we assume that the observable
is integrable with respect to this state. 

In the upper panel of Fig. \ref{SM125}, we see a simulated sample of the time series of the potential energy, of a particle in a Lennard-Jones potential (the details of the simulation are similar to Fig. \ref{SM25}). The time series exhibits long periods where \eq{|V(x)|<\epsilon}, and \eq{\epsilon} is some lower cutoff that can be as small as we wish, and short periods of high energy (in absolute value). In the lower panel of Fig. \ref{SM125}, we see that the probability distribution \eq{g(\tau)} of the durations (\eq{\tau}s) of the events where the energy is low, which correspond to events where the particle has strayed far away from the potential minimum, has a power-law shape \eq{g(\tau)\propto\tau^{-3/2}}, at large \eq{\tau} (as seen from the fitting function, in an orange line), like in free Brownian motion, as expected. Hence the process is recurrent, but the mean-return time is infinite. In this example, we used \eq{\epsilon=10^{-5}}.
{We verified the Mittag-Leffler distribution of several observables, including the potential energy, using various numerical simulations, whose results are presented in our previous paper, Ref. \cite{aghion2019non}.}

\begin{figure}[t]		
\centering
\includegraphics[width=1.0\columnwidth]{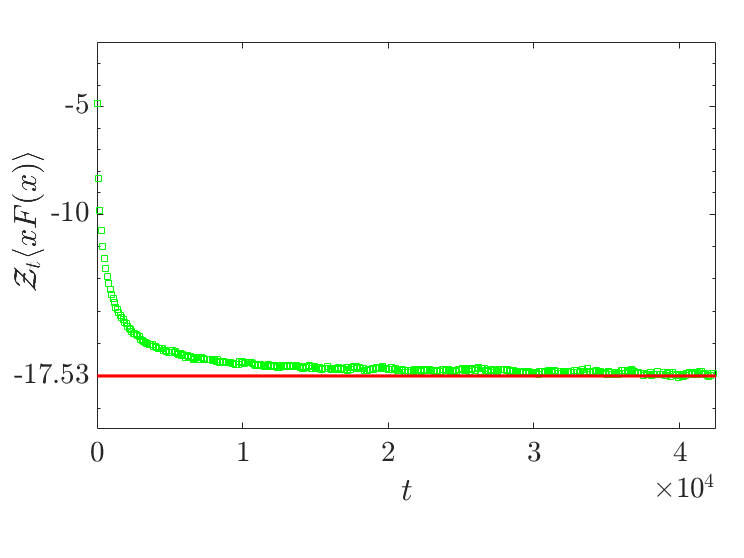} 
\caption[Virial Theorem 1]{{\footnotesize{The value of \eq{\mathcal{Z}_t\langle xF(x)\rangle}, obtained from simulation results based on numerical integration over the Langevin Eq. \eqref{eq08}, where \eq{F(x)=-V'(x)}, and \eq{V(x)=(-25 x^2 + x^4/2)\exp(-x^2/20)/125} (green circles), approaches at increasing times to the theoretical value given by Eq. \eqref{eqV05}. Here \eq{\mathcal{Z}_t=2\sqrt{\pi D t}}, since the potential is symmetric. \eq{\gamma=1}, and \eq{D=k_BT=0.5}.}}}
\label{FigVirialTheoremFig11}
\end{figure}

 \begin{figure}[t]
\includegraphics[width=1.0\linewidth]{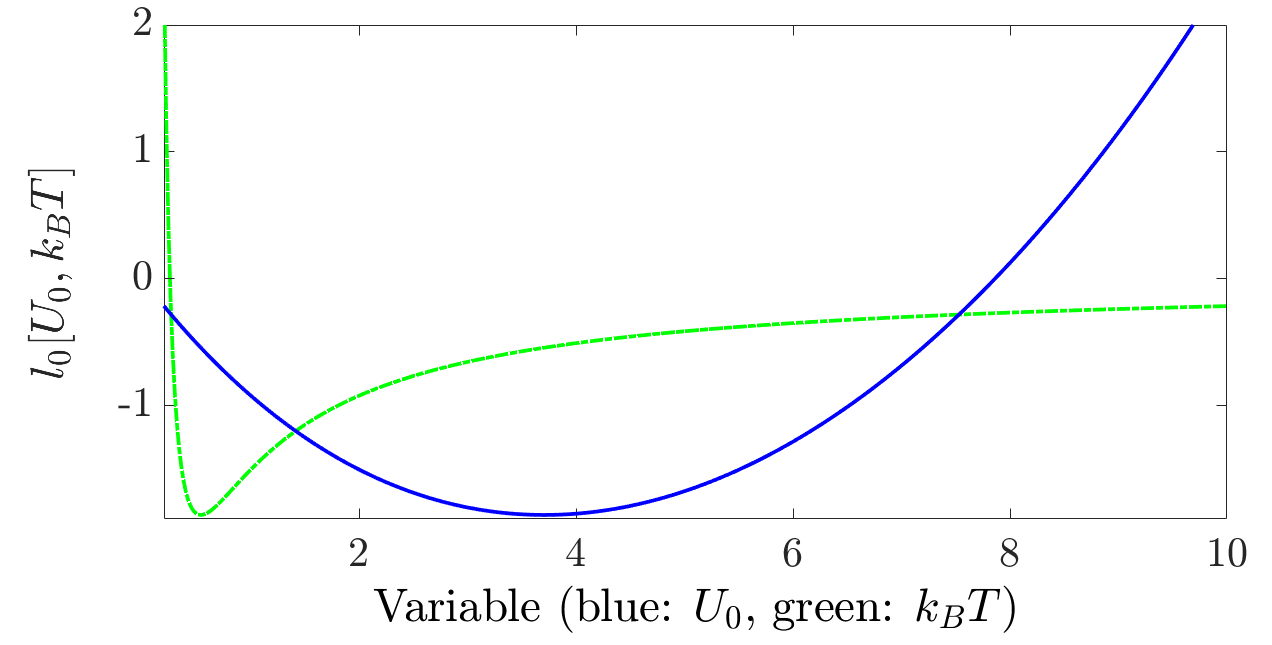} 
\caption{{\footnotesize{(Color online). Various values of \eq{l_0}, obtained for the potential \eq{U(x,U_0)=U_0\left[(-x/5)^4-(x/5)^2\right]\exp\left[(-x/5)^2\right]}, which change sign for various values of the amplitude \eq{U_0} (blue line), at fixed temperature: \eq{k_BT=1}, as well as for a fixed \eq{U_0=2}, and varying \eq{T} (green line).}}}
\label{Figl0}
\end{figure}

\section{Virial Theorem}
\label{VirialTheorem125}

The virial theorem addresses the mean of the observable $x F(x)$, where  
$F(x)=-V'(x)$.
 binding potentials, treated with standard thermodynamics, yield $\langle{x F(x)}\rangle=-k_B T$. In our case \cite{aghion2019non},  using the non-normalized Boltzmann state we find, by integration by parts 
\begin{align}
    \langle x F(x) \rangle &= {k_B T \over \mathcal{Z}_t}  \int_0 ^\infty x {\partial \over \partial x} \left( e^{ - V(x)/k_B T } - 1 \right) {\rm d} x=\nonumber\\ 
    & -{k_B T \over \mathcal{Z}_t  } \int_0 ^\infty \left(  e^{ - V(x)/k_B T} -1 \right)  {\rm d} x , 
\label{eqV052}
\end{align}
where we used our convention $\exp[ - V(\infty) /k_B T ] = 1$.
Now, using \eq{\ell_0=\int_0^\infty[\exp(-V(x)/k_BT)-1]\Intd x} (introduced in Sec. \ref{SubSecEigen}), we get 
\begin{equation}
\lim_{t\rightarrow\infty}\mathcal{Z}_t\langle x F(x) \rangle \rightarrow - k_B T {\ell_0}. 
\label{eqV05}
\end{equation}
 The ratio $\ell_0 /\mathcal{Z}_t$ distinguishes Eq. \eqref{eqV05} from the standard thermal virial theorem, where the ratio \eq{-\langle xF(x)\rangle/k_BT} at equilibrium is unity. Note that \EQ{\ell_0=-2{B}_2,}{SecondVirialCoefficient} where \eq{{B}_2} is called the second virial coefficient \cite{chandler1988introduction}. For two-sided, symmetric potentials  \eq{\langle xF(x)\rangle=-2k_BT\ell_0/\mathcal{Z}_t} (but as mentioned, also  \eq{\mathcal{Z}_t\rightarrow2\mathcal{Z}_t}). 
The constant \eq{l_0} points to a surprising link between the virial theorem and the corrections to the uniform approximation, studied in Sec. \ref{SubSecEigen} (particularly, Eq. \eqref{OuterCorrection1} for large \eq{x}), which means that by measuring the shape of the tails of the diffusing particle packet, one can, at least in principle, obtain knowledge about the force in the system, even though in the large \eq{x} region it is effectively zero.  
Interestingly, notice that \eq{\ell_0} can change sign, for various potentials, which is also very different from standard thermodynamics. 
 Fig. \ref{FigVirialTheoremFig11} shows the approach of the simulated value of \eq{\mathcal{Z}_t\langle xF(x)\rangle} where \eq{F(x)=-V'(x)}, and \eq{V(x)=(-25 x^2 + x^4/2)\exp(-x^2/20)/125} at increasing times (green circles), to the theoretical limit, Eq.  \eqref{eqV05} (red line) with \eq{k_BT=\gamma=0.5}, where the value is negative. This result was obtained from the overdamped Langevin Eq. \eqref{eq08}.   In Fig. \ref{Figl0}, we show the various values of \eq{l_0}, obtained for the potential \eq{U(x,U_0)=U_0\left[(x/5)^4-(x/5)^2\right]\exp\left[(-x/5)^2\right]}, for various values of the amplitude \eq{U_0} (blue line), at fixed temperature: \eq{k_BT=1}, as well as for a fixed \eq{U_0=2}, and varying \eq{T} (green line). In both those cases we used \eq{\gamma=1}.


\subsection{A Correction term for the mean-squared displacement, and the underdamped Langevin process} 
\label{NextToLeadingOrderBehaviorOfTheMSD}

The result of the previous subsection provides us with a nice example that demonstrates that the next-to-leading order correction term, derived for the uniform approximation of \eq{P_t(x)} in Sec. \ref{SubSecEigen}, has also some relevance to thermodynamics. This link is made by looking at the correction term to the linear behavior, in time, of the mean-squared displacement \eq{\langle x^2\rangle} of the system, which we now derive using an elementary calculation, and then explain it in terms of the virial theorem. 

\begin{figure}[t]
\includegraphics[width=0.8\linewidth]{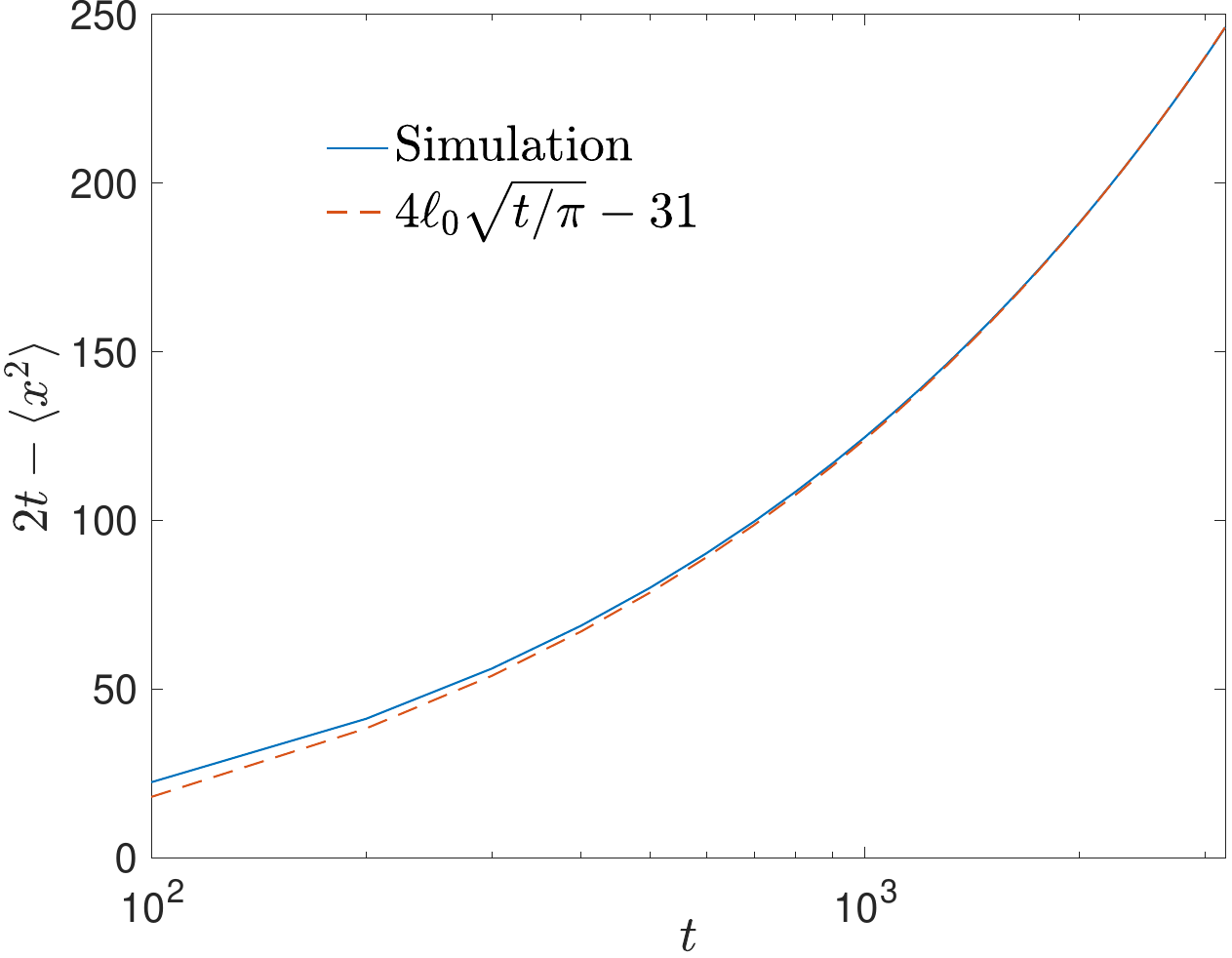} 
\caption{{\footnotesize{(Color online) Numerical results (blue line), corresponding to overdamped Langevin dynamics with the same potential used for all the figures in Sec. \ref{SubSecEigen}, with \eq{k_BT=\gamma=1}), which confirm the validity of the correction term to the leading-order, linear, behavior of the second moment in time. The fitting curve is shown in orange-dashed line, and it corresponds to the \eq{\sqrt{t}} term in Eq. \eqref{eqA5}, plus an additional constant coming from higher-order correction terms, which was obtained numerically.}}}
\label{FigmsdisplaceFigureFromDavid}
\end{figure}

Using Eq. \eqref{OuterCorrection1}, we get  
\begin{align} 
\centering
&\langle x^2\rangle\approx\int_0^\infty \frac{x^2}{\sqrt{\pi Dt}}e^{-x^2/(4Dt)}\Intd x\nonumber\\ 
&\qquad\qquad -\int_0^\infty \frac{x^2e^{-x^2/(4Dt)}}{\sqrt{\pi Dt}}\left[ 
     \frac{\ell_0 x}{2Dt}\right]\Intd x \nonumber\\ 
&=2Dt-4{\ell_0}\sqrt{Dt/\pi}, 
\label{eqA5}
\end{align} 
where we used the fact that at large \eq{x}, \eq{V(x)\approx0}. In the above derivation, we considered the integration limits to be zero and infinity, although Eq. \eqref{OuterCorrection1} is exact only in the large \eq{x} limit, since the contribution to the mean-squared displacement from the small \eq{x} regime is negligibly small at the long-time limit.  Fig. \ref{FigmsdisplaceFigureFromDavid} shows numerical results corresponding to overdamped Langevin dynamics with the same potential used for all the figures in Sec. \ref{SubSecEigen}, with \eq{k_BT=\gamma=1}, which confirm the validity of the correction term to the leading-order,  linear, behavior of the second moment in time. The figure also shows an additional constant coming from higher-order correction terms, which was obtained numerically.  Note that in the case of a two-sided system, there might be additional correction terms to the mean-squared displacement, of order \eq{\sqrt{t}}, if the initial position of the particle is not located at the origin. The reason is that, here, the correction terms to \eq{P_t(x)} might differ from Sec. \ref{SubSecEigen}. \\ 

\begin{figure}[t]
\includegraphics[width=1.0\linewidth]{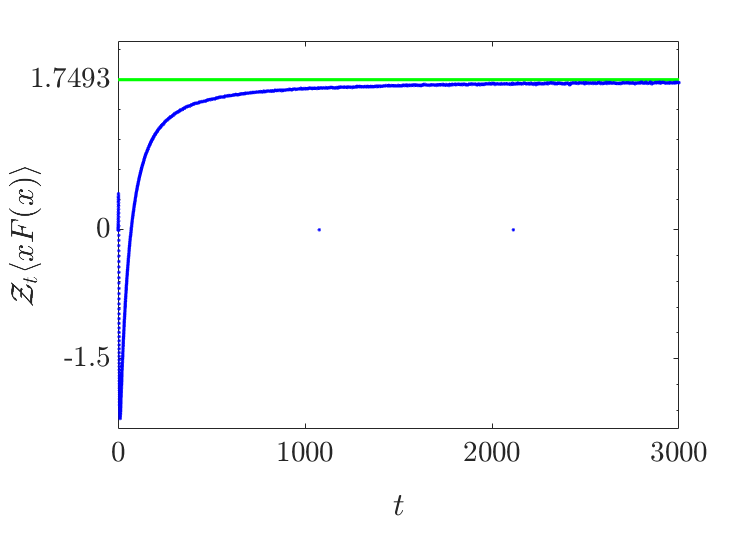} 
\caption{{\footnotesize{(Color online) The approach of ~\eq{\mathcal{Z}_t\langle xF(x)\rangle} obtained from simulation results corresponding to the underdamped Langevin process with the two-sided potential Eq. \eqref{eq091}, with \eq{k_BT=0.8} and \eq{\gamma=0.5}, at increasing times (blue circles) to the asymptotic theoretical value Eq. \eqref{eqV05} (green line).}}}
\label{meanxFx}
\end{figure} 

To understand the connection between Eq. \eqref{eqA5} and the virial theorem, we need to extend our analysis to the phase space and consider both the particle's position \eq{x}, and it's velocity \eq{v}. Namely, in what follows, instead of Eq. \eqref{eq08} we now use the underdamped Langevin equation, with zero-mean,  white Gaussian noise;
 $m \partial_t v = F(x) - \gamma v + \eta(t)$. In this process, which also obeys
the fluctuation-dissipation relation, \eq{\gamma>0}, and we include
 the acceleration term
according to Newton's second law, where $m$ is the particle's mass. The analysis below will yield the same results regardless if \eq{V(x)} is a one-sided or two-sided potential, given that the process starts at \eq{x(t=0)=0}. 
Consider  the identity
\begin{equation}
m \partial_t \langle x v \rangle = m \langle v^2 \rangle + m \langle x \partial_t v \rangle. 
\label{eqV01}
\end{equation}
 Since  the velocity is thermalised
 $m \langle v^2 \rangle = k_B T$, and since $\langle x \eta(t) \rangle = 0$ we get
\begin{equation}
m \partial_t \langle x v \rangle = k_B T + \langle x F(x) \rangle - \gamma \langle x v  \rangle,
\label{eqV02}
\end{equation}
where clearly
\begin{equation}
\langle x v \rangle = {\partial_t \langle x^2 \rangle \over 2} .
\label{eqV02a}
\end{equation}
Furthermore, we assume that in the long-time limit 
\begin{equation}
\langle x^2 \rangle \sim 2 D t +  D_\beta t^\beta + ...,
\label{eqV03}
\end{equation}
where the second term is small compared with the first.
It is then clear that  $|m \partial_t \langle x v \rangle| \ll|\gamma \langle x v \rangle|$
and hence 
\begin{equation}
\gamma \langle x v \rangle = k_B T + \langle x F(x) \rangle .
\label{eqV04}
\end{equation}
For equilibrium situations, i.e. for  binding potentials like the Harmonic oscillator,
$\langle x v \rangle=0$ since the marginal position density is described
by the Boltzmann distribution, and  Maxwell-distribution for the velocities. This means that in thermal equilibrium, 
the velocities are not correlated with the spatial  position of the particle, 
since in the single particle  Hamiltonian
the kinetic energy is separated from the potential  energy. 
For the case under study here, the correlation $\langle x v \rangle $ is not strictly zero. 

 
Coming back to Eq. 
(\ref{eqV04}), we see that the term $\langle x F(x) \rangle$ decays like $t^{-1/2}$,
since $\mathcal{Z}_t \propto t^{1/2}$. Using the leading order term  $\langle x v \rangle= \partial _t \langle x^2 \rangle/2\sim  D$,  from  Eq. 
(\ref{eqV04}) we get in the very long-time limit  $k_B T - \gamma D=0$, recovering  the Einstein relation. 
Hence we need to consider the sub-leading terms. It is easy to see that since $\langle x F(x) \rangle \propto t^{-1/2}$ we must have $\beta=1/2$. 
Using Eqs. (\ref{eqV02a},\ref{eqV03},\ref{eqV04})
we find
\begin{equation}
\langle x F(x) \rangle = { D_{1/2} \gamma \over 4 t^{1/2}},
\label{eqV06}
\end{equation}
and from 
Eq. (\ref{eqV05})
\begin{equation}
D_{1/2} = - { 4 \sqrt{D} \ell_0 \over  \sqrt{\pi}}. 
\label{eqV07}
\end{equation}
It follows that $D_{1/2}$ is determined by the potential energy via the
length-scale $\ell_0$, and for a given $D$ it is independent of the mass of the particle. Using Eq. (\ref{eqV07}) in Eq. \eqref{eqV03}, we  therefore recover the first and second leading order terms in \eq{\langle x^2\rangle},  obtained in Eq. \eqref{eqA5}. 
Fig. \ref{meanxFx} shows the approach of the mean \eq{\langle xF(x)\rangle} obtained from simulation results, using the underdamped Langevin equation with the potential Eq. \eqref{eq091}, to the asymptotic value Eq. \eqref{eqV05}.

 We remark that the observable $x v$ is non-integrable with respect to 
the non-normalized Boltzmann-Gibbs  state. In phase space, we speculate that the non-normalized state
is $\exp[ - H /k_B T]/\mathcal{Z}_t$ where the Hamiltonian $H= p^2/2 m + V(x)$ and $p= m v$ as usual. 
The only change here is that
$\mathcal{Z}_t$ is now  $\sqrt{ \pi D t}\sqrt{ 2 \pi m k_B T}$, where the factor 
$\sqrt{ 2 \pi m k_B T}$ stems from the Maxwell distribution.
This expression describes the bulk fluctuations of the packet of particles in phase space, 
while in the far tails the
correlation between
$x$ and $v$ builds up. The full analysis of the phase-space infinite-density remains out of the scope of this manuscript. 

\section{Non-normalizable Boltzmann-Gibbs states in $d$-dimensions}
\label{dDimenssional} 

So far,  we have treated only one-dimensional processes. However, as we mentioned in the introduction, the issue of the non-normalizability of the Boltzmann factor raised by Fermi was in the context of three-dimensional motion, under the influence of a Coulomb potential \cite{fermi1924wahrscheinlichkeit}. We now show that 
the non-normalizable Boltzmann-Gibbs state is found also in $d$-dimensions, when the external  potential is isotropic, and it decays at least as rapidly as \eq{1/r}, at large distances. One should keep in mind here that in $2$-dimensions, in the absence of any potential field, Polya's theorem states that a Brownian particle still returns to its origin with probability \eq{1} (and the mean first return time is infinite), but in any dimension \eq{d>2}, this is no longer the case, if the system size is infinite. This holds also  in the presence of Coulomb-like potentials. Still, as we now show, the Boltzmann infinite density is valid.  

We begin our analysis in the absence of any force. 
The radial motion of a Brownian particle in \eq{d}-dimensions, in the space defined by the orthogonal directions \eq{\chi_1,\chi_2,...\chi_d}, is described by Bessel process \cite{kubo2012statistical}  \EQ{\dot{r}=D\frac{d-1}{r}+\sqrt{2D}\Gamma(t),}{BesselProcess}
where \eq{r=\sqrt{\chi_1^2+\chi_2^2+...\chi_d^2}.} Accordingly, the radial Fokker-Planck equation describing the expansion of the probability density \eq{W_t(r)} in \eq{d} dimensions is \cite{kubo2012statistical} \eq{{\partial_t}W_t(r)=D\left[((d-1)/r)\partial_r+\partial^2_r\right]W_t(r),} where   \eq{\int_0^\infty W_t(r)1r^{d-1}\Intd r=1} and \eq{c(d)} is a constant which rises from the integration over all the angular degrees of freedom of the \eq{d}-dimensional Laplacian. Here we assumed that the initial distribution of the particles was also isotropic around the origin. Substituting \EQ{W_t(r)=P_t(r)r^{1-d}/c(d),}{FPEWP} yields  
\EQ{\frac{\partial}{\partial t}P_t(r)=D\left[\frac{\partial}{\partial r}\frac{{1-d}}{r}+\frac{\partial^2}{\partial r^2}\right]P_t(r),}{FPIndDimensionsForPtr}
where \eq{\int_0^\infty P_t(r)\Intd r=1}. The solution to this equation, for various boundary conditions, is found e.g. in \cite{bray2000random,martin2011first,medalion2016size}. In two dimensions, for example, starting from a ring-shaped initial distribution $P_0(r,\phi)=\delta(r-r_0)\Theta(0<\phi<2\pi)/(2\pi r_0),$  with a reflecting boundary condition at \eq{r=0} (see details in \cite{bray2000random}), at time \eq{t} we find  $P_t(r)\approx ({r}/{2Dt})\exp\left(-({r^2+r_0^2})/{4Dt}\right)I_{0}\left({rr_0}/{2Dt}\right),$ where~\eq{I_\nu(\cdot)} is the Bessel function with index \eq{\nu} (and here \eq{\nu=0}). In \eq{d}-dimensions, starting from a uniform probability distribution on a \eq{d}-dimensional sphere of radius \eq{r_0}, and a reflecting boundary condition at \eq{r=0} \cite{bray2000random}: 
\EQ{P_t(r)= \frac{r_0}{2Dt}\left(\frac{r}{r_0}\right)^{d/2}\exp\left(-\frac{r^2+r_0^2}{4Dt}\right)I_{d/2-1}\left(\frac{rr_0}{2Dt}\right).}{TwoDimensional0} 
Below, we add to the system an asymptotically flat potential field, with a repelling  part on the origin. In that case, the right-hand side of Eq. \eqref{TwoDimensional0} will describe the shape of the density in the large \eq{r} regime, where \eq{V(r) \sim0}. One can easily verify, for example, that when \eq{d=1} and \eq{r_0\ll t^{1/2}}, but $r\sim \mathcal{O}(t^{1/2})$, \eq{P_t(r)} in Eq. \eqref{TwoDimensional0} is $\sim \frac{1}{\sqrt{\pi Dt}}\exp(-r^2/(4Dt))$ (as is well known \cite{bray2000random}).

\subsection{The infinite-invariant density}

Consider an isotropic, radially dependent potential \eq{V(r)}, such that \eq{V(0)={\infty}} and \eq{V(\infty)={0}}, which falls-off at least as rapidly as \eq{1/r} at large distances (in the same spirit as the potentials we investigated in the unidimensional case). Now, {the radial dynamics  is described by
 
\EQ{\dot{r}=D\frac{(d-1)}{r}-\frac{V'(r)}{\gamma}+\sqrt{2D}\Gamma(t).}{BesselWithIsotropicPotential} As in the unidimensional Langevin equation, Eq. \eqref{eq08}, here \eq{D=k_BT/\gamma} and \eq{k_T,\gamma>0}. The corresponding radial Fokker-Planck equation is  
\EQ{{\frac{\partial}{\partial t}}W_t(r)=D\left[\frac{d-1}{r}\frac{\partial}{\partial r}+r^{1-d}	\frac{\partial}{\partial r}r^{d-1}\frac{V'(r)}{k_BT}+\frac{\partial^2}{\partial r^2}\right]W_t(r).}{} 
Here, and in what follows, we assume that the initial particle distribution is narrow, and isotropic. 
After repeating the substitution from Eq. \eqref{FPEWP}, this yields  
\EQ{\frac{\partial}{\partial t}P_t(r)=D\left[\frac{\partial}{\partial r}\frac{{1-d}}{r}+\frac{\partial}{\partial r}\frac{V'(r)}{k_BT} +\frac{\partial^2}{\partial r^2}\right]P_t(r).}{Bray} 
To solve Eq. \eqref{Bray} to leading order, in the long-time limit, we use the ansatz \eq{P_t(r)\approx \frac{r_0}{2Dt}\left(\frac{r}{r_0}\right)^{d/2}\exp\left(-\frac{r^2+r_0^2}{4Dt}\right)I_{d/2-1}\left(\frac{rr_0}{2Dt}\right)\mathcal{I}(r),} where \eq{\mathcal{I}(r)} is some general function of \eq{r}. Note that this approach is similar to that employed in the unidimensional case, in Sec. \ref{NnBGState}. Plugging this ansatz in Eq. \eqref{Bray}, we obtain the uniform approximation  
\EQ{P_t(r)\approx \frac{r_0}{2Dt}\left(\frac{r}{r_0}\right)^{d/2}e^{-\frac{r^2+r_0^2}{4Dt}}I_{d/2-1}\left(\frac{rr_0}{2Dt}\right)e^{-\frac{V(r)}{k_BT}},}{TwoDimensional}
 for long \eq{t}. From the uniform approximation, Eq. \eqref{TwoDimensional}, using the asymptotic shape of the Bessel function in the limit \eq{t\rightarrow\infty}, since  
$I_{d/2-1}\left(\frac{rr_0}{2Dt}\right)\approx ({rr_0})^{d/2-1}/\left((4{Dt})^{d/2-1}\Gamma(d/2)\right)$, and \eq{\exp[-(r^2+r_0^2)/(4Dt)]\rightarrow1}, we find      \EQ{\lim_{t\rightarrow\infty}{\mathcal{Z}_t}r^{1-d}P_t(r)\rightarrow\exp\left(-V(r)/k_BT\right),}{TwoDimensional1} where \eq{{\mathcal{Z}}_t=2^{d/2-1}\Gamma(d/2)(2Dt)^{d/2}}.  
Importantly, from this relation we again  see that, in the long time limit, the non-normalizable solution is independent of \eq{r_0}. In one dimension, from this result we recover Eq. \eqref{UniformApproximation}. 

  \begin{figure}[t]
\includegraphics[width=1.0\linewidth]{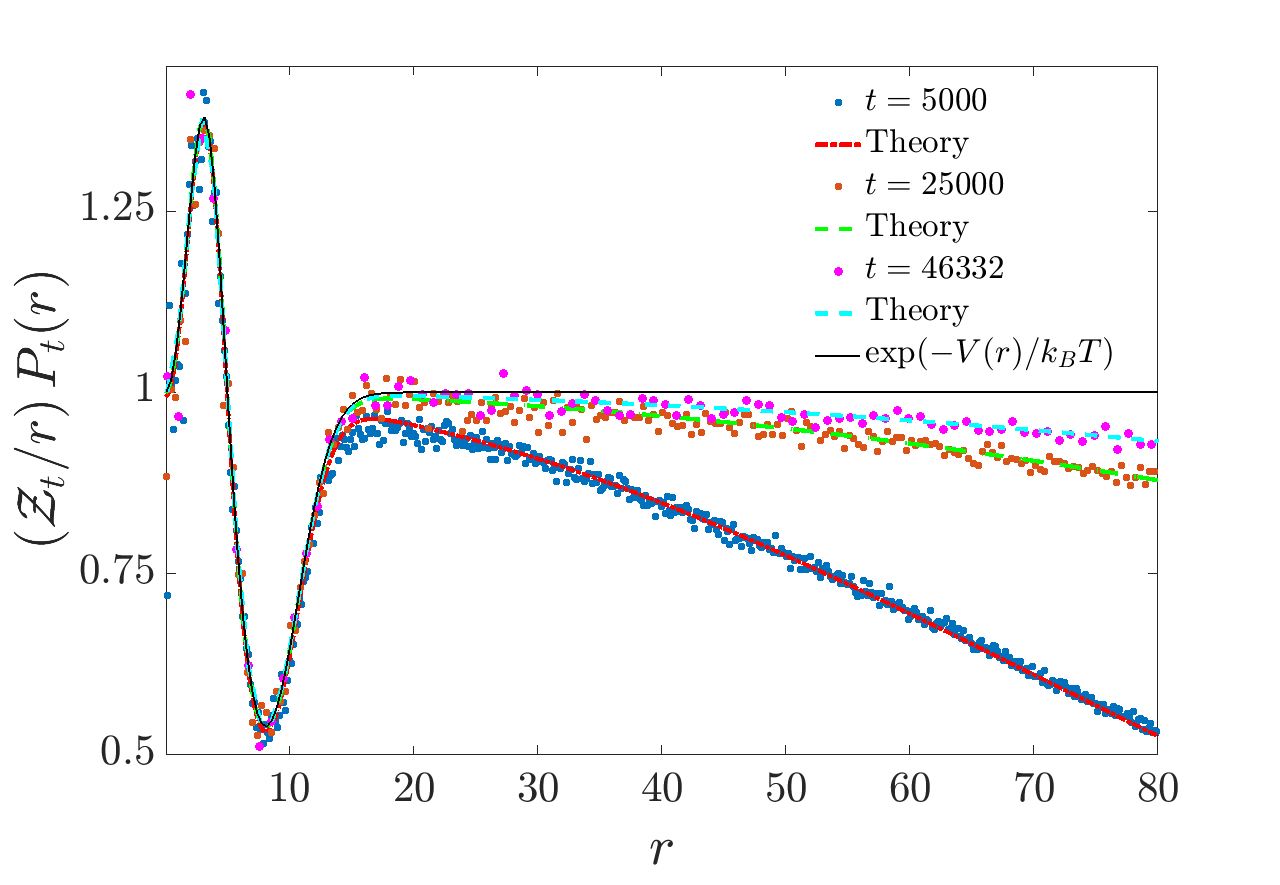} 
    \caption[Non-normalizable Boltzmann state in two dimensions]{{\footnotesize{(Color online) Simulation results for a two-dimensional Langevin process,  with \eq{V(r)} in Eq. \eqref{IsotropicPotential} (\eq{\gamma=1}, and \eq{k_BT=D=0.5}), {compared with}  theory. Colored symbols:  \eq{(\mathcal{Z}_t/r)P_t(r)}, where \eq{r=\sqrt{x^2+y^2}}, \eq{\mathcal{Z}_t=2D t} (see Eq. \eqref{TwoDimensional1}) and \eq{P_t(r)} is the radial distribution, obtained from the simulation at times \eq{t= 5000, 25000,46322} (blue circles, orange squares and magenta circles, respectively). The corresponding theoretical curves obtained from Eq. \eqref{TwoDimensional} with \eq{d=2} (multiplied by \eq{2Dt/r}) are the red dash-dot line and a green dashed line and cyan dashed-line. The non-normalizable Boltzmann factor \eq{\exp(-V(r)/k_BT)}, appears in solid black line. Here, the initial distribution of the \eq{10^6} particles was uniform around a ring of radius \eq{r_0=5}.}}}
\label{FigTwoDimension}
\end{figure}
 
  Fig. \ref{FigTwoDimension} shows excellent agreement between the simulation results of a two-dimensional Langevin process, and the theory corresponding to Eqs. (\ref{TwoDimensional},\ref{TwoDimensional1}).  The simulation method used an Euler-Mayurama integration scheme over the two Langevin equations $\dot{x}=-V'(r)\cos(\theta)+\sqrt{2D}\Gamma_1(t)$ and $\dot{y}=-V'(r)\sin(\theta)+\sqrt{2D}\Gamma_2(t)$, where \EQ{V(r)=\left[(r/5)^4-(r/5)^2\right]\exp[-(r/5)^2].}{IsotropicPotential} 
 ${\Gamma_1(t)}$ and ${\Gamma_2(t)}$ represent two independent, zero-mean and ${\delta}$-correlated Gaussian white noise terms. At \eq{t=0}, the particles were uniformly distributed around a ring of radius \eq{r_0=5}.  
   Fig. \ref{FigThreeDimensions} shows simulation results of a three-dimensional Langevin process, with the Coulomb-type potential \EQ{V(r)=(2/r)^{12}-0.5/r.}{CouloumbType} Here \eq{r} is defined in spherical coordinates. The repulsive part of the potentials, which falls-off as rapidly as \eq{1/r^{12}}, was added to regularize the interactions near the origin in the simulation (this is technically simpler to realize numerically than putting a hard spherical wall around some \eq{r\ll1}). This model mimics a field created by a finite-sized charge, which repels  the observed particle at short distances. The figure shows excellent match between the simulated radial PDF \eq{P_t(r)}, multiplied by \eq{\mathcal{Z}_t} (defined in Eq. \eqref{TwoDimensional1}) at times \eq{t=5000,50000,100000} (green stars, blue circles and red diamonds, respectively),  and the corresponding uniform approximation, Eq. \eqref{TwoDimensional} with \eq{d=3} (magenta, purple and brown lines, respectively). At increasing times, the simulation results approach the non-normalizable Boltzmann state \eq{\exp(-V(r)/k_BT)} (black line), via Eq. \eqref{TwoDimensional1}, as expected, confirming the existence of the infinite-invariant density also in three dimensions. Here, \eq{D=k_BT=0.05}. {In this system the minimum of the field is on \eq{r_{\textrm{min}} \approx 2.84}, hence in Fig. 16 we see a peak at this value. Also notice that the depth of the well is \eq{V(r_{\textrm{min}}) \approx -0.16}, hence \eq{V(r_{\textrm{min}}) / k_BT \approx -3.22}. 
Thus we are dealing here with a trap that is not too deep, this allows the escape of the particles  on a finite observation time. 
For a deeper well, we will need to wait for even longer observation time to find the infinite density. }
 
 \begin{figure}[t]
\includegraphics[width=1.0\linewidth]{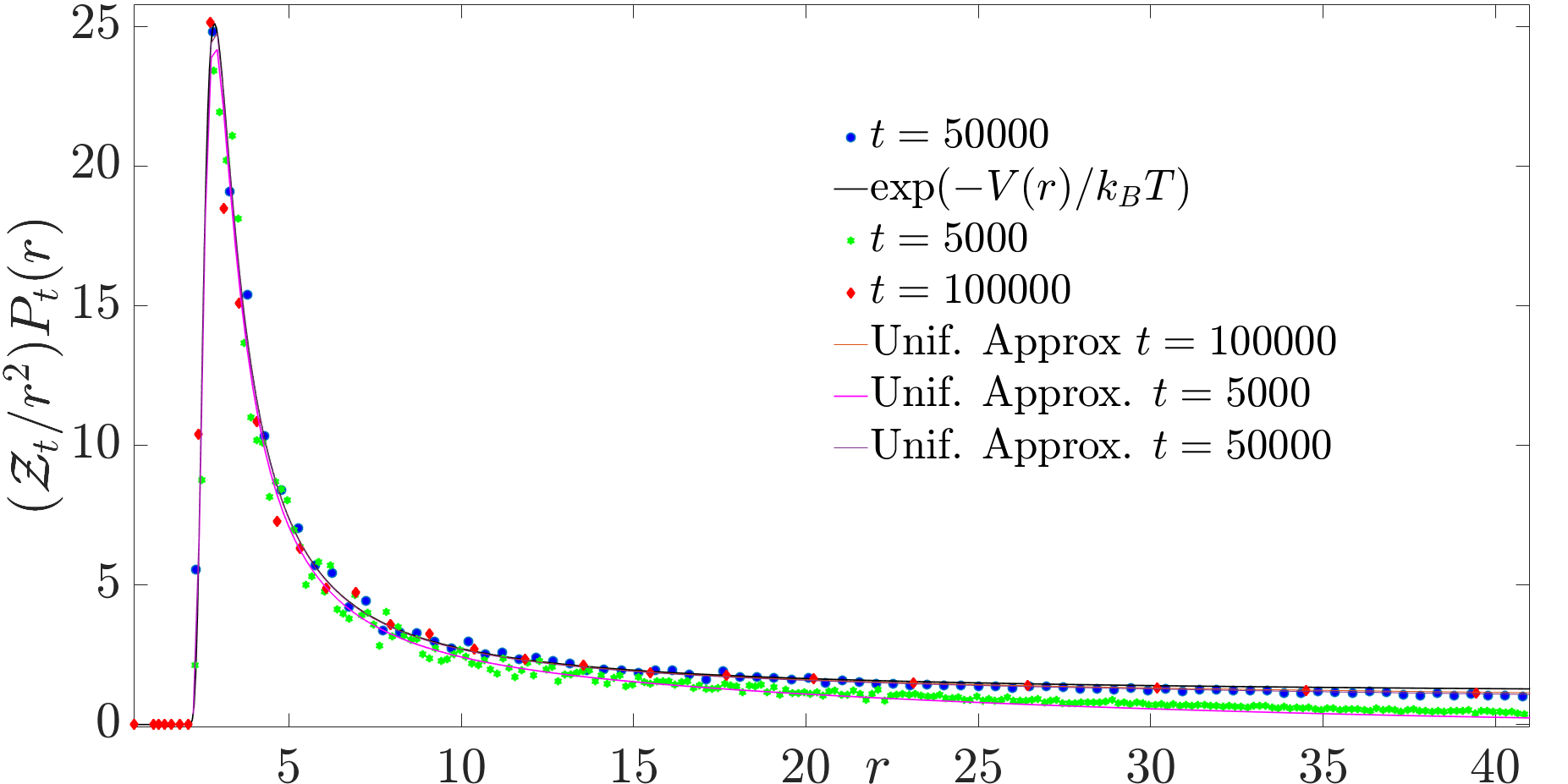} 
    \caption[Non-normalizable Boltzmann state in three dimensions, with Coulomb-type interaction]{{\footnotesize{(Color online) Simulation results of a three-dimensional Langevin process, with the Coulomb-type potential Eq. \eqref{CouloumbType}. The figure shows excellent match between the simulated radial PDF \eq{P_t(r)} (here we used spherical coordinates), multiplied by \eq{\mathcal{Z}_t/r^2} (and \eq{{\mathcal{Z}}_t} is defined below Eq. \eqref{TwoDimensional1}) at times \eq{t=5000,50000,100000} (green stars, blue circles and red diamonds, respectively),  and the corresponding uniform approximation , Eq. \eqref{TwoDimensional} with \eq{d=3} (magenta, purple and brown lines, respectively). At increasing times, the simulation results approach to the non-normalizable Boltzmann state \eq{\exp(-V(r)/k_BT)} (black line), via Eq. \eqref{TwoDimensional1}, as expected, confirming the existence of the infinite-invariant density also in three dimensions. Here, \eq{D=k_BT=0.05}, and we used \eq{10^7} particles.}}}
\label{FigThreeDimensions}
\end{figure}

\subsection{Ergodicity of time-weighted observables, in $d$-dimensions}
\label{ErgodicityIndDimensions} 

As mentioned above, Brownian motion in $d$-dimensions is non-recurrent. In this section, we propose a new approach for evaluating time and ensemble averages of observables integrable with respect to the infinite-density, Eq. \eqref{TwoDimensional1}, and show that this method leads to a Birkhoff-like equality between the two means, which is valid in any dimension \eq{d\geq1}. 

Let \eq{\mathcal{O}(r)} be an integrable observable, with respect to \eq{r^{d-1}\exp\left(-V(r)/k_BT\right)}, in Eq. \eqref{TwoDimensional1}, e.g., \eq{\mathcal{O}(r)=\Theta(r_a<r(t)<r_b)}. Here, 
\eq{\Theta(r_a<r(t)<r_b)=1} while the particle's trajectory passes inside the $d$-dimensional shell with inner radius \eq{r_a>0}, and outer radius \eq{r_b<\infty}, and zero otherwise. We define the ensemble mean of the \textit{weighted} time-average as 
\begin{align}
    &\left\langle\overline{\mathcal{Z}_t\mathcal{O}[r(t)]}\right\rangle\equiv\left\langle\frac{1}{t}\int_0^t \mathcal{Z}_{\tilde{t}}\mathcal{O}[r(\tilde{t})]\Intd \tilde{t}\right\rangle\nonumber\\ 
    &=\frac{1}{t}\int_0^t \mathcal{Z}_{\tilde{t}}\left\langle\mathcal{O}(r)\right\rangle_{\tilde{t}}\Intd \tilde{t}=\frac{1}{t}\int_0^t \Intd \tilde{t}\mathcal{Z}_{\tilde{t}}\int_0^\infty\mathcal{O}(r)P_{\tilde{t}}(r)\Intd r. 
    \label{WeightedTA1}
\end{align} 
\begin{figure}[t] 
\includegraphics[width=1.0\linewidth]{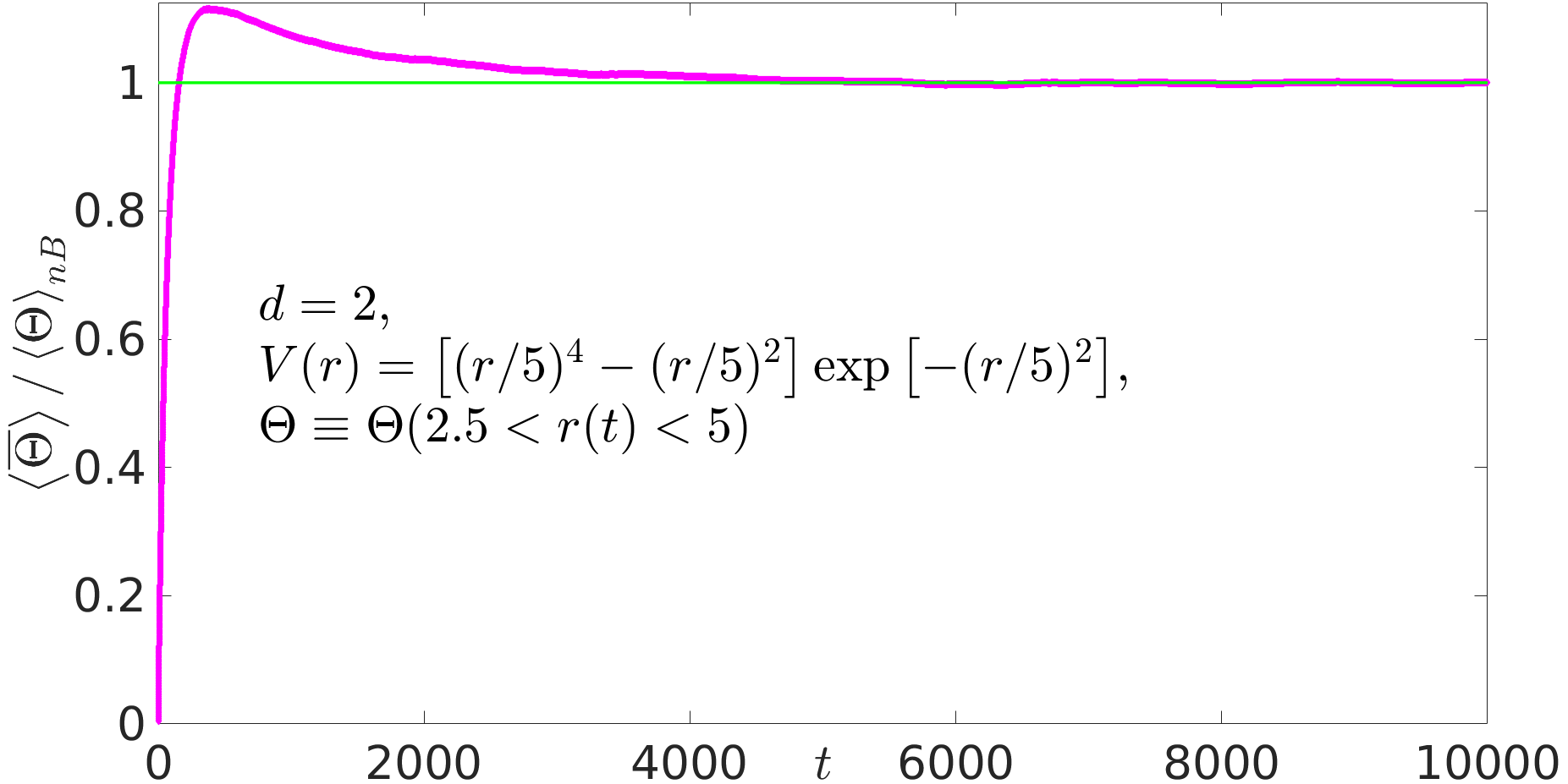} 
\caption{{\footnotesize{(Color online) The ratio between the weighted time average Eq. \eqref{WeightedTA1}, measured in a two-dimensional Langevin simulation, and the ensemble average with respect to the non-normalizable Boltzmann state \eq{} \eq{\langle\Theta\rangle_{nB}} of the indicator function \eq{\Theta(2.5<r(t)<5)}, converges to unity at increasing times (simulation results in magenta line, the green line on \eq{1} serves as a guide to the eye). This validates Eq. \eqref{ErgodicityIndDimensions1}. The details of the simulation  are similar to those in Fig. \ref{FigTwoDimension} (so \eq{\langle\Theta\rangle_{nB}\approx11.7076}).}}}
\label{ErgodicityTwoDimensionsWeightedMean}
\end{figure} 
Using Eq. \eqref{TwoDimensional1}, in the limit  \eq{t\rightarrow\infty}, Eq. \eqref{WeightedTA1} leads to  
\begin{align}
    \left\langle\overline{\mathcal{Z}_t\mathcal{O}[r(t)]}\right\rangle&\rightarrow\frac{1}{t}\int_0^t \Intd \tilde{t}\mathcal{Z}_{\tilde{t}}\left[\frac{1}{\mathcal{Z}_{\tilde{t}}}\right]\int_0^\infty\mathcal{O}(r)r^{d-1}e^{-V(r)/k_BT}\Intd r, 
    \nonumber\\
    &=\int_0^\infty\mathcal{O}(r)r^{d-1}e^{-V(r)/k_BT}\Intd r. 
    \label{WeightedTA2}
\end{align} 
We denote \eq{\langle\mathcal{O}(r)\rangle_{nB}\equiv \int_0^\infty\mathcal{O}(r)r^{d-1}e^{-V(r)/k_BT}\Intd r,} and then Eq. \eqref{WeightedTA2} means that when \eq{t\rightarrow\infty}, \eq{\left\langle\overline{\mathcal{Z}_t\mathcal{O}[r(t)]}\right\rangle\rightarrow\langle\mathcal{O}(r)\rangle_{nB}}. We now relate this weighted time average to an ensemble average, performed at time \eq{t}. Clearly,  
\begin{align}
    \mathcal{Z}_t\left\langle{\mathcal{O}(r)}\right\rangle_t&=\mathcal{Z}_t{\int_{0}^\infty}\mathcal{O}(r)P_t(r)\Intd r\nonumber\\ &\rightarrow \langle\mathcal{O}(r)\rangle_{nB}.  
    \label{WeightedEA}
\end{align} 
 Therefore, 
\begin{align}
    \lim_{t\rightarrow\infty}\frac{\left\langle\overline{\mathcal{Z}_t\mathcal{O}[r(t)]}\right\rangle}{\left\langle{\mathcal{Z}_t\mathcal{O}(r)}\right\rangle_t}\rightarrow 1. 
    \label{ErgodicityIndDimensions1}
\end{align} 
Eq. \eqref{ErgodicityIndDimensions1} is a generalized form of Birkhoff's theorem. 
Note that in one dimension, this equation constitutes an alternative to Eq. \eqref{eqTA12}, which was derived for non-weighted time-averages. 
Here, like in Sec. \ref{SecTimeAndEnsembleMeans}, both the ensemble-average, and the weighted time-average are estimated in the long-time limit from the non-normalizable ($d$-dimensional) Boltzmann-Gibbs state, Eq. \eqref{TwoDimensional1}, hence Eq. \eqref{ErgodicityIndDimensions1} constitutes further extension of infinite-ergodic theory (in the sense of the ratio between time and ensemble means).  Fig. \ref{ErgodicityTwoDimensionsWeightedMean} shows the ratio between the weighted time average Eq. \eqref{WeightedTA1}, measured in a two-dimensional Langevin simulation, and the ensemble average with respect to the non-normalizable Boltzmann state  \eq{\langle\Theta\rangle_{nB}} of the indicator function \eq{\Theta(2.5<r(t)<5)}. This ratio converges to unity at increasing times, validating Eq. \eqref{ErgodicityIndDimensions1}. The details of the simulation  are similar to Fig. \ref{FigTwoDimension}. 
Fig. \ref{ErgodicityThreeDimensionsWeightedMean} shows a similar ratio as in Fig. \ref{ErgodicityTwoDimensionsWeightedMean}, but this time it is measured in a three-dimensional Langevin simulation, for \eq{\Theta(2.8<r(t)<3)} (blue line). The details of this  simulation  are similar to Fig. \ref{FigThreeDimensions}, and the ratio again converges to unity at increasing times. 
\begin{figure}[t] 
\includegraphics[width=1.0\linewidth]{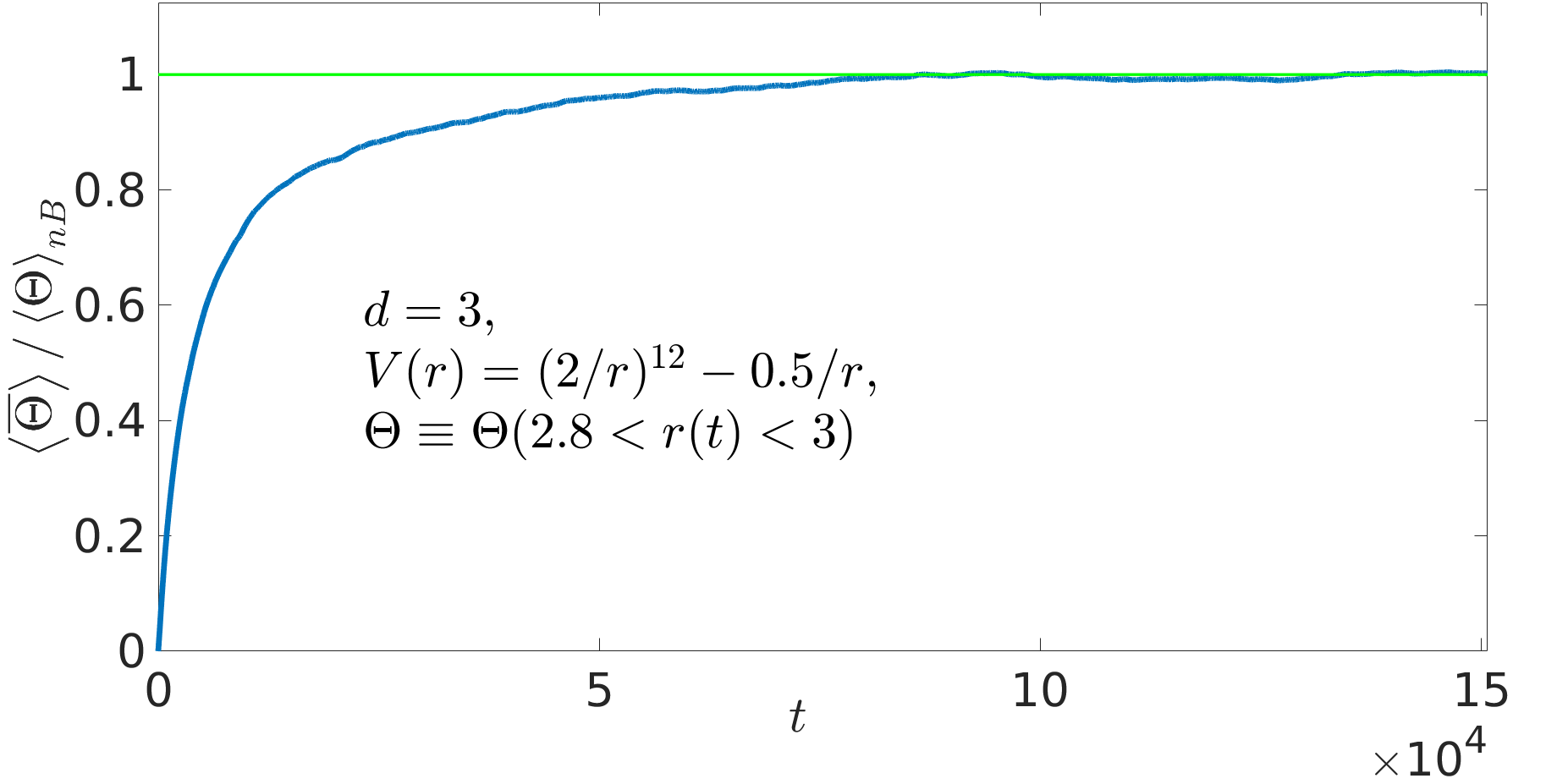} 
\caption{{\footnotesize{(Color online) The ratio between the weighted time average Eq. \eqref{WeightedTA1}, measured in a three-dimensional Langevin simulation, and the ensemble average with respect to the non-normalizable Boltzmann state  \eq{\langle\Theta\rangle_{nB}} of the indicator function \eq{\Theta(2.8<r(t)<3)}, converges to unity at increasing times (simulation results in blue line, the green line on \eq{1} serves as a guide to the eye). This validates Eq. \eqref{ErgodicityIndDimensions1}. The details of the simulation  are similar to Fig. \ref{FigThreeDimensions} (so \eq{\langle\Theta\rangle_{nB}\approx41.7144}), except that here we used \eq{10^5} particles.}}}
\label{ErgodicityThreeDimensionsWeightedMean}
\end{figure} 


\section{Summary of our main results} 
\label{SectionReview}
 
 In this manuscript, which extends our work in Ref. \cite{aghion2019non}, we have shown that the spatial shape  of a diffusing particle packet, inside an asymptotically flat potential, converges at long times to a non-normalizable Boltzmann state in any dimension, \eq{d\geq1}. This state, which is treated mathematically as an infinite-invariant density \cite{aaronson1997introduction}, takes the place of the standard Boltzmann distribution, which gives the equilibrium state in systems with strong confinement, in the sense that we can use it to obtain time and ensemble averages of integrable observables. 
 We have mainly focused on one-dimensional systems, which obey the Aaronson-Darlin-Kac theorem, and here we also showed that the non-normalizable Boltzmann state gives the  entropy-energy relation, and the virial theorem. We studied the {emergence of} the latter in detail in one dimension, and it would be interesting to study it further also in higher dimensions in a future work. 
 
 We have obtained the non-normalizable Boltzmann state in one dimension using three different techniques: via physical scaling assumptions (Sec. \ref{NnBGState}), using the entropy maximization principle (Sec. \ref{SecEntropyEnergy}), and via a rigorous eigenfunction expansion method (Sec. \ref{SubSecEigen}). The last of these also yielded terms which describe the sub-leading order behavior of the probability density function, which decay more rapidly with time. Though the analysis based on an eigenfunction expansion in \eq{d>1} dimension is left for future work, here we  showed that by attaching a proper weight function to integrable observables, the ratio between weighted time averages and ensemble averages converges to unity (see Eq. \eqref{ErgodicityIndDimensions1}). The distribution of the weighted time average is an open question for future research. 
 
 The main results of this manuscript are  the uniformly valid approximation for the one-dimensional probability density $P_t(x)$ including the first-order correction, Eq. (\ref{UNIFORM}); the relationship between the mean-squared position and the virial theorem, expressed in  Eq. (\ref{eqA5}) and Eqs. (\ref{eqV03} and \ref{eqV07}); the leading order probability density in arbitrary dimensions, Eq. (\ref{TwoDimensional1}); and lastly, the ergodicity of time-weighted observables, expressed in Eq. (\ref{ErgodicityIndDimensions1}). 
 

\section{Discussion} 
\label{Discussion}

 Infinite ergodic theory can be applied to many thermodynamic systems, as the main condition is that  the fluctuation-dissipation theorem holds. One must distinguish however
between recurrent, and non-recurrent 
processes, since only in the latter the Aaronson-Darlin-Kac theorem holds. Physically, the key point is that we can identify easily important observables that are integrable with respect to the infinite density, and the fluctuation-dissipation theorem guarantees that this infinite density is the non-normalizable Boltzmann factor.   Ryabov, et al. \cite{martin2018diffusion,Holubeck2019living} considered a  different, though related, setup in one dimension, with an unstable potential that 
does not allow for the return of the particle to its starting point.
Also here the partition function diverges, but again a certain aspect of Boltzmann equilibrium remains. Hence, our work suggests that further aspects of ergodicity should be studied in this setup, perhaps in the spirit of the evaluation of weighted-time-averaged observables, and it also indicates that future studies in that direction could be  interesting in other classes of potentials as well. 
Our work also encourages the investigation of these problems in non-Markovian settings, and in the presence of many-body interactions{, and since we have assumed that the fluctuation-dissipation relation holds, it will be interesting to examine if this assumption can be relaxed and infinite-ergodic theory can be studied also e.g. in the framework of active particles,  as in \cite{romanczuk2012active,hoell2019multi}}. 

While in this manuscript we considered the fluctuations of time averages, and in particular the fluctuations of energy, the whole framework of stochastic thermodynamics could in principle be investigated. 
 For example, it would be interesting to
explore  the fluctuations of the rate of entropy  production, and the work and heat exchange between the particles and the heat bath.
It should be noted however that our theory gives rise to both extensions of stochastic thermodynamics, for systems with a non-normalizable Boltzmann-Gibbs state, but also the connection between fluctuations (diffusivity) and thermodynamics. This is seen  in the virial correction to the diffusion law (Sec. \ref{VirialTheorem125}). In the current theory, one cannot separate diffusion 
from non-normalized Boltzmann-Gibbs states, as was demonstrated
in the extremum principle studied in Sec. \ref{SecEntropyEnergy}. 
\\ 

\textbf{Acknowledgement:} The support of Israel Science Foundation grant $1898/17$ is acknowledged.

\appendix 

\bibliographystyle{aipnum4-1}
\bibliography{./bibliography2}

\end{document}